\documentclass[11pt,epsf]{article}
\usepackage{hyperref}
\usepackage{subfigure}
\usepackage{amsmath}
\usepackage{graphics}
\usepackage{amssymb}
\usepackage{epsf}
\usepackage{rotating}
\usepackage{pstricks}

\def\eps{\epsilon}
\def\a{\alpha}
\def\b{\beta}

\def\g{\gamma}

\def\si{\sigma}
\def\lag{\langle}
\def\rag{\rangle}

\newcommand{\bea}{\begin{eqnarray}}
\newcommand{\eea}{\end{eqnarray}}

\newcommand{\psl}{p \! \! \!  /}
\newcommand{\qsl}{q \! \! \!  /}
\newcommand{\lsl}{l \! \! \!  /}

\newcommand{\beqa}{\begin{eqnarray}}
\newcommand{\eeqa}{\end{eqnarray}}
\def\beq{\begin{equation}}
\def\eeq{\end{equation}}
\def\nn{\nonumber}

\begin{document}

\begin{center}
\vspace{.5cm}

{\bf\large The Trace Anomaly and the Gravitational Coupling of an Anomalous $U(1)$}

\vspace{1.5cm}
{\bf\large Roberta Armillis, Claudio Corian\`{o}, Luigi Delle Rose and Luigi Manni
\footnote{roberta.armillis@le.infn.it, claudio.coriano@le.infn.it, luigi.dellerose@le.infn.it, luigi.manni@le.infn.it}}
\vspace{1cm}

{\it  Dipartimento di Fisica, Universit\`{a} del Salento \\
and  INFN Sezione di Lecce,  Via Arnesano 73100 Lecce, Italy}\\
\vspace{.5cm}

\begin{abstract}
We extend a previous computation of the $TJJ$ correlator, involving the energy-momentum tensor of an abelian gauge theory and two vector currents ($J\equiv J_V$), to the case of mixed axial-vector/vector currents ($J_A$). The study is performed in analogy to the case of the $AVV$ vertex for the chiral anomaly. We derive the general structure of the anomalous Ward identities and provide explicit tests of their consistency using Dimensional Reduction. Mixed massive correlators of the form $TJ_V J_A$ are shown to vanish both by Ward identities and by C-invariance. The result is characterized by  the appearance of massless scalar degrees of freedom in the coupling of chiral and vector theories to gravity, affecting both the soft and the ultraviolet region of the vertex. This is in agreement with previous studies of the effective action of gauge and conformal anomalies in QED and QCD.  
\end{abstract}
\end{center}
\newpage

\section{Introduction}
In a previous work we have presented a complete computation of the off-shell graviton-photon-photon vertex for an abelian
gauge theory, which is derived from the correlator of the energy-momentum tensor $(T)$ with two vector currents $(J)$
 (the $TJJ$ correlator) \cite{Giannotti:2008cv, Armillis:2009pq}. Previous studies of this correlator include those of \cite{Berends:1975ah, Drummond:1978hh, Drummond:1979pp, Dolgov:1981nw}, which were limited to the QED case, while, surprisingly, there has not been any previous attempt to discuss the structure of more general vertices, such the $TJ_AJ_A$ or $TJ_V J_A$ correlators,  carrying one insertion of the energy momentum tensor and of one or more chiral currents. 
 
 These correlators appear in the expression of the 1 particle irreducible (1PI) effective action which describes the interaction of gravity with the fields of a chiral theory, such as the 
 Standard Model, and contribute, to leading order in the gauge coupling expansion, to the radiative breaking of scale invariance. In turn, this is the prominent perturbative feature of the trace anomaly, which appears to be generated by specific pole terms, as we are going to 
 elaborate below.
  
  Correlators of this type can potentially carry mixed anomalies. Specifically, this can be a trace anomaly, due to the insertion of an energy momentum tensor, in combination with a chiral anomaly, due to the presence of axial-vector currents. This anomaly mixing, in principle, is expected to be present both in the case that we investigate - involving one or two axial-vector currents - and in higher point functions. In the latter case they may involve a larger number of axial-vector gauge currents, such as the $TJ_A J_AJ_A$ vertex and many others, which are divergent by power-counting, as one can easily figure out, and contribute to higher  perturbative orders.
  
  As in the case of the diagram responsible for the chiral anomaly (the axial-vector/vector/vector, or $AVV$ diagram), also in the 
  case under analysis one of the crucial points relies on the derivation of the correct Ward identities which allow to define this trilinear vertex consistently.  This point requires some care, due to the formal manipulations involved in the handling of the functional integral and to the presence of mass corrections. In the massless case, instead, the computation of this correlator can be formally related to the vector case (the $TJJ$ case) of \cite{Giannotti:2008cv, Armillis:2009pq} by a naive manipulation of the chiral projectors in the loops. Our investigation addresses all these points in some detail, offering a general approach that can be applied to the realistic case of the Standard Model. In this respect, the study of the gravitational coupling of a chiral abelian theory (with one anomalous $U(1)$) contains all the issues that appear in of the fermion sector of the non-abelian case.  
 \subsection{The anomalous effective action}
As we have mentioned above, one of the key features of the trace anomaly is the appearance in the 1PI effective action of dynamical massless poles which mediate the anomalous interaction 
\cite{Giannotti:2008cv, Armillis:2009pq}. 
The story of massless poles in anomaly-mediated interactions, obviously, is not new, and goes back to Dolgov and Zakharov \cite{Dolgov:1971ri}, in their analysis of the chiral anomaly. The nonlocal "$1/\square$" structure of the effective anomalous interaction, due to the pole term in the correlator, is, in fact, a distinctive feature of the diagrammatic expansion of these effective theories. These can be made local at the cost of introducing two pseudoscar (auxiliary) fields \cite{Coriano:2008pg}. In the case of conformal anomalies, the identification of similar massless poles and their interpretation has been addressed recently in \cite{Giannotti:2008cv}, and in \cite{Armillis:2009pq}, by direct computations. These singularities, as discussed in these works, affect both the infrared and the ultraviolet region of the anomaly diagrams, as we will illustrate in the next sections. These features, present in the QED and QCD cases, are naturally shared by an anomalous abelian theory when it gets coupled to gravity.

The possible physical implications of this behaviour of the effective action have been discussed in \cite{Mottola:2010gp}, and for this reason similar analysis in the complete Standard Model and for other correlators (such as the $TTT$ vertex ) are underway. 

\subsection{Aspects of the computation}
Coming to other features of our computation, it should be remarked that a direct derivation from first principles of correlators with axial-vector/vector currents and energy momentum insertions, in general, runs into difficulties. This is due to the appearance of commutators of the energy momentum tensor with the chiral current, situation that we will try to avoid.  

As in the vector-like case, we will provide explicit expressions of all the form factors appearing in the correlator, for a simple theory.
We have selected an abelian model with two vector/axial-vector currents and a single massive fermion. 
One important point 
that we intend to stress is that the local (gauge) or global nature of the two currents, in the example that we provide, 
is not relevant for the conclusions and the goals of this analysis, being the two gauge fields to which the two currents couple just classical background fields. For this reason, our investigation is essentially the
search of the correct conditions for defining anomalous correlators of the form $TJ_VJ_A$ and $TJ_AJ_A$ (with a single insertion of $T_{\mu\nu}$). The approach 
is the exact analogous of the one followed in the investigation of the AVV graph of the chiral anomaly, and in principle could be generalized to more complex correlators. Unfortunately, however, the explicit test of the Ward identities containing higher point functions 
becomes increasingly difficult in perturbation theory. 

 Another remark concerns the use of Dimensional 
Reduction (DRED) with a 4-dimensional $\gamma_5$ \cite{Siegel:1979wq} in our analysis. Typically, in these types of studies, it is necessary
at each step to check the consistency of the perturbative result against the constraints posed by the anomalous Ward identities. Our results, which are more complex than in a previous analysis of the $TJJ$ vertex, indeed satisfy these conditions. It has also been checked that Dimensional Regularization (DR) and DRED give the same expression for the $TJ_AJ_A$ vertex, while they differ in the case of the $TJ_V J_A$ vertex by infinite contributions. In this second case, as we are going to show, both the condition of charge conjugation invariance (C-invariance) and the Ward identity extracted from the functional integral imply that this specific vertex is required to vanish identically for any fermion mass. 

\section{The Lagrangian and the off-shell effective action}
To establish notations, here we will briefly summarize our conventions. The diagrammatic contributions will be presented both in the usual $V/A$ (vector/axial-vector) form, with Dirac spinors, and in the $L/R$ (Left-Right) form, using chiral fermions. We will include mass effects in the fermion loops and we will keep all the external lines off their mass-shell in order to establish the most general form of the corresponding effective action. 

We consider a theory with a Dirac fermion $\psi$ and two abelian gauge bosons, namely $V$ and $A$, described by the Lagrangian
\bea
\mathcal L_0 = -\frac{1}{4}F_{V \,\mu \nu}F_V^{\mu \nu} -\frac{1}{4}F_{A \,\mu \nu}F_A^{\mu \nu} + \bar \psi\gamma^\mu(i \partial_\mu + g V_\mu + g \gamma^5 A_\mu)\psi - m \bar \psi \psi,
\label{lagr}
\eea
where the fermion couples to the two gauge bosons with, respectively, a vector and an axial-vector interaction. In our conventions, the axial-vector gauge boson is denoted by $A$, while the vector one is denoted by $V$. 
The axial current will be denoted $J_A^{\mu} = \bar \psi \gamma^{\mu} \gamma_5 \psi$, and sometimes we will be using a suffix ``5" to emphasize its axial-vector character. For instance $\Pi_{55}\def \Pi_{AA}$ will denote the axial-axial
two-point function while $\Pi\equiv \Pi_{VV}$ will denote the corresponding two-point function of the vector case. In the derivation of the Ward identities which will be discussed below, the gauge fields will be considered as external background fields both in the $V/A$ and in the $L/R$ formulation. This theory couples to gravity in the weak gravitational field limit  
via the energy momentum tensor of (\ref{lagr}). 

In particular, the corresponding effective action will be formally defined as the sum of 

1) the tree-level action given by (\ref{lagr}) \\

\beq
\mathcal{S}_0= \int d^4 x \mathcal{ L}_0
\eeq
and
2) the trilinear interactions $T J_A J_V, T J_V J_V$ and $T J_A J_A$. These extra graphs appear as leading 
corrections to the effective action, which is defined as

\beq
\mathcal S_{anom}\equiv \langle\Gamma_{AA} h AA\rangle + \langle\Gamma_{VA} h V A\rangle +  \langle\Gamma_{VV}h V A\rangle
\eeq
with 
\beq
\langle\Gamma_{h AA}\rangle\equiv \int d^4 z\, d^4 x\, d^4 y \,\Gamma_{A
A}^{\mu\nu\alpha\beta} h_{\mu\nu}(z) \,A_{\alpha}(x)\, A_{\beta}(y)\,
\eeq
and similarly for all the other terms. The field $h_{\mu\nu}$ denotes the linearized fluctuations of the metric around a flat background
\beq
g_{\mu\nu}= \eta_{\mu\nu} +\kappa h_{\mu\nu}, \qquad\qquad \kappa=\sqrt{16 \pi G_N}
\eeq
with $G_N$ being the 4-dimensional Newton's constant.

One of the principal goals of our investigation is to provide a correct definition of $\mathcal{S}_{anom}$ by deriving the essential Ward identities of the anomalous correlators. At the same time we will show, as in a previous case study for QED, that the effective action is characterized by massless anomaly poles. The extraction of these singularities, in our case, 
 is not based on dispersion theory as in \cite{Giannotti:2008cv} but the results are obviously equivalent to the dispersive treatment \cite{Armillis:2009pq} in the massless case, with a generalization for massive fermions.
 
 \subsection{Symmetries and the energy momentum tensor}
The Lagrangian in (\ref{lagr}) remains invariant under the local vector gauge transformation $U(1)_V$
\bea
\psi &\rightarrow& e^{i g \, \alpha(x)}\psi, \\
\bar \psi &\rightarrow& \bar \psi e^{-i g \, \alpha(x)}, \\
V^{\mu} &\rightarrow& V^{\mu} + \partial^{\mu} \alpha(x),
\eea
which implies the conservation of the vector current $J_V^{\mu} \equiv J^{\mu} = \bar \psi \gamma^{\mu} \psi$. If the fermion mass is zero the Lagrangian is also invariant under a local axial-vector gauge transformation $U(1)_A$
\bea
\psi &\rightarrow& e^{i g \, \beta(x)\gamma_5}\psi, \\
\bar \psi &\rightarrow& \bar \psi e^{i g \, \beta(x) \gamma_5}, \\
A^{\mu} &\rightarrow& A^{\mu} + \partial^{\mu} \beta(x),
\eea
implying the conservation of the axial-vector current $J_A$. Obviously, this is explicitly broken by the contributions of massive fermions
\bea
\partial_{\mu} J_A^{\mu} = 2 i m \, \bar \psi \gamma_5 \psi.
\eea
The energy-momentum tensor consists of four contributions: the free fermion part $T_f$, the fermion-boson interaction parts $T_{i_V}$ and $T_{i_A}$, due to the interactions of the axial and vector gauge fields with the fermions, and the gauge term $T_g$ which are given by
\beq
T^{\mu\nu}_{f} = -i \bar\psi \gamma^{(\mu}\!\!
\stackrel{\leftrightarrow}{\partial}\!^{\nu)}\psi + g^{\mu\nu}
(i \bar\psi \gamma^{\lambda}\!\!\stackrel{\leftrightarrow}{\partial}\!\!_{\lambda}\psi
- m\bar\psi\psi),
\label{tfermionic}
\eeq
\beq
T^{\mu\nu}_{i_V} = -\, g J^{(\mu}V^{\nu)} + g g^{\mu\nu}J^{\lambda}V_{\lambda}\,,
\eeq
\beq
T^{\mu\nu}_{i_A} = -\, g J_A^{(\mu}A^{\nu)} + g g^{\mu\nu}J_A^{\lambda}A_{\lambda}\,,
\label{taxial}
\eeq
and
\beq
T^{\mu\nu}_{g} = F_V^{\mu\lambda}F^{\nu}_{V\,\lambda} - \frac{1}{4} g^{\mu\nu}
F_V^{\lambda\rho}F_{V\,\lambda\rho} + F_A^{\mu\lambda}F^{\nu}_{A\,\lambda} - \frac{1}{4} g^{\mu\nu}
F_A^{\lambda\rho}F_{A\,\lambda\rho}.
\label{tphoton}
\eeq
The complete energy-momentum tensor is
\bea
T^{\mu\nu} = T^{\mu\nu}_{f}+T^{\mu\nu}_{i_V}+T^{\mu\nu}_{i_A}+ T^{\mu\nu}_{g},
\eea
which couples to gravity with a linearized term of the form $h_{\mu\nu}T^{\mu\nu}$. The Lagrangian (\ref{lagr}) can be rewritten in the chiral basis decomposing the fields in terms of their left-handed and right-handed components by using the chirality projectors
\bea
P_L = \frac{1- \gamma_5}{2}, \qquad \qquad P_R = \frac{1 + \gamma_5}{2}.
\eea
We define the chiral fermion fields as
\bea
\psi_L = P_L \psi, \qquad \qquad \psi_R = P_R \psi
\eea
and the left and right gauge fields, $A_L$ and $A_R$, as
\bea
A_L^{\mu} &=& V^{\mu} - A^{\mu}, \\
A_R^{\mu} &=& V^{\mu} + A^{\mu},
\eea
so that the Lagrangian takes the form
\bea
\mathcal L =
-\frac{1}{4}F_{L \,\mu \nu} \, F_L^{\mu \nu} - \frac{1}{4}F_{R \,\mu \nu}\, F_R^{\mu \nu}
+ \bar \psi_L \gamma_\mu(i \, \partial^\mu + g A_L^{\mu})\,  \psi_L
+ \bar \psi_R \gamma_\mu(i \, \partial^\mu + g A_R^{\mu}) \, \psi_R
\label{lchiral}
\eea
when the mass term has been set to vanish. The energy momentum is separated into the various chiral contributions
\bea
T^{\mu\nu}_{f,L} &=& -i \bar\psi \gamma^{(\mu}\!\!
\stackrel{\leftrightarrow}{\partial}\!^{\nu)} \, P_L \psi + g^{\mu\nu}
i \bar\psi \gamma^{\lambda}\!\!\stackrel{\leftrightarrow}{\partial}\!\!_{\lambda}\, P_L \psi, \\
T^{\mu\nu}_{f,R} &=& -i \bar\psi \gamma^{(\mu}\!\!
\stackrel{\leftrightarrow}{\partial}\!^{\nu)} \, P_R \psi + g^{\mu\nu}
i \bar\psi \gamma^{\lambda}\!\!\stackrel{\leftrightarrow}{\partial}\!\!_{\lambda}\, P_R \psi, \\
T^{\mu\nu}_{i,L} &=& -\, g \,( J_L^{(\mu}A_L^{\nu)} - g^{\mu\nu}J_L^{\lambda}A_{L {\lambda}})\,, \\
T^{\mu\nu}_{i,R} &=& -\, g \, ( J_R^{(\mu}A_R^{\nu)} - g^{\mu\nu}J_R^{\lambda}A_{R\lambda})\,,
\eea
with
\bea
J_L^{\mu}(x) &=& \bar\psi(x)\gamma^\mu P_L \psi(x), \\
J_R^{\mu}(x) &=& \bar\psi(x)\gamma^\mu P_R \psi(x).
\eea
Notice that the Lagrangian in (\ref{lchiral}) is invariant under the chiral transformation $U(1)_L \times U(1)_R$.

\subsection{Perturbative expansion of the axial-vector contributions}
The analysis of the vector-like contributions, i.e. of the $\langle T J  J \rangle$ correlator,  has been performed in great detail in
\cite{Armillis:2009pq}. For this reason we will consider, at this point, a vanishing vector contribution $(V\to 0)$ in the defining Lagrangian (\ref{lagr}) and we will focus our discussion at the moment on its axial part. A relation between the vector and axial contributions will be worked out in the later sections, where we will show that mixed vector-axial vector correlators vanish for any nonzero $m$. We will also show how to relate pure vector like to axial vector like contributions, as indicated below
in Eq. \ref{GammaSplit}.  

To extract the one-loop contributions to the $\langle T  J_A J_A \rangle$ correlator in  the perturbative expansion and identify 
those due to the conformal anomaly, it is sufficient to consider only the partial energy-momentum tensor $T_p$ given by the Dirac and the interaction term in eqs.~(\ref{tfermionic}) and (\ref{taxial})
\bea
T_p^{\mu\nu} = T^{\mu\nu}_{f}+T^{\mu\nu}_{iA},
\eea
while the gauge term in eq.(\ref{tphoton}) is only responsible, to second order ($g^2$), of two non-amputated diagrams removed from the perturbative expansion of the effective action. We also recall that the conservation of the energy momentum tensor can be reformulated as a partial conservation equation
\beq
\partial_\nu T_p^{\mu\nu}= -\partial_\nu T_{A g }^{ \mu\nu},
\label{classical}
\eeq
with
\beq
T_{A g}^{\mu\nu}\equiv  F_A^{\mu\lambda}F^{\nu}_{A\,\lambda} - \frac{1}{4} g^{\mu\nu}
F_A^{\lambda\rho}F_{A\,\lambda\rho}.
\eeq

Using diffeomorphism invariance  one can derive formally a quantum relation similar to (\ref{classical}), which takes the form 
\beq
\partial_\nu \langle T_p^{\mu\nu} \rangle_A= g \, F_A^{\mu\lambda} \, \langle J_{A\lambda} \rangle_A.
\label{naiveWI}
\eeq
This relation is the analogue - for the axial case - of the relation identified in \cite{Giannotti:2008cv}, 
which allows to extract the momentum conservation Ward identity in the case of the $TJJ$ (for vector currents).
In (\ref{naiveWI}) the functional average of $T_p^{\mu\nu}$ is now defined as 
\beqa
\langle T_p^{\mu\nu}(z)\rangle_A &\equiv& \int D\psi D\bar{\psi} \,\,T^{\mu\nu}_p (z) \,\,e^{i \int d^4 x \, \mathcal{L}_k(\psi) + i g \int d^4 x \, J_A\cdot A(x) }
\eeqa

with 
\beq
\mathcal{L}_k(\psi)\equiv \bar{\psi} i \gamma^\mu \partial_\mu \psi
\eeq
being the kinetic fermion Lagrangian in flat spacetime, and we will denote by $\mathcal{S}_k(\psi)$ the corresponding action. Notice that equation (\ref{naiveWI}) can be naively thought as the quantum counterpart of the non-homogeneous equation 
\beq
\partial_\nu T_p^{\mu\nu} = g \, F_A^{\mu\lambda} \, J_{A\lambda} 
\label{naiveWIs}
\eeq
satisfied by $T_p^{\mu\nu}$. Here the axial vector field $A$ is taken as a background. A rigorous derivation of this relation requires the use of invariance under diffeomorphism of the generating functional of the 
full theory (expressed in terms of $g_{\mu\nu}$ and a $A_\mu$) and an expansion around flat space, as can be checked.  

The conservation equation (\ref{naiveWI}) is relevant for the extraction of one of the Ward identities necessary to define the correlator.  Notice that the expectation value of $T_p$ in the background of the gauge field $A$ is the generating functional of the correlation functions that we need. These are obtained by an expansion through second order in the external field $A$. The relevant terms in this expansion are explicitly given by
\beq
\langle T_p^{\mu\nu}(z)\rangle_A =
\frac{(i g)^2} {2!} \, \langle T_{f}^{\mu\nu}(z) \, (J_A\cdot A) \, (J_A\cdot A) \rangle +
i g \, \langle T_{i A}^{\mu\nu}(z) \, (J_A\cdot A)\rangle + ... \, ,
\eeq
with  $(J_A\cdot A)\equiv \int d^4 x \, J_A\cdot A(x)$.

The corresponding diagrams are extracted via two functional derivatives respect to the background field $A$
and are given by
\beq
\Gamma_{AA}^{\mu\nu\alpha\beta} (z; x, y) \equiv \frac{ \delta^2 \lag T_p^{\mu\nu} (z) \rag_A}
{\delta A_{\alpha}(x)\delta A_{\beta}(y)} \bigg\vert_{A=0}
= V_{55}^{\mu\nu\alpha\beta} (z; x, y)+ W_{55}^{\mu\nu\alpha\beta} (z; x, y),
\eeq
where
\beq
 V_{55}^{\mu\nu\alpha\beta} (z; x, y)=(i \, g )^2 \, \lag T_{f}^{\mu\nu} (z) J_A^{\alpha} (x) J_A^{\beta} (y) \rag_{A=0},
 \label{V55c}
\eeq
 and  
\beqa
W_{55}^{\mu\nu\alpha\beta}(z; x, y) &=& (i \, g ) \, \frac{ \delta^2 \lag T_{iA}^{\mu\nu} (z) (J_A\cdot A) \rag}
{\delta A_{\alpha}(x)\delta A_{\beta}(y)} \bigg\vert_{A=0} \nonumber \\
&=& \delta^4(x-z)g^{\alpha (\mu} \Pi_{AA}^{\nu )\beta}(z, y)
+ \delta^4 (y-z)g^{\beta(\mu} \Pi_{AA}^{\nu )\alpha}(z, x)
- g^{\mu\nu}[\delta^4(x-z) + \delta^4(y-z) ]\Pi_{AA}^{\alpha\beta}(x, y),\nn \\
\label{tbubble}
\eea
is a second term expressed in terms of  the correlator of two axial currents
\bea
\Pi^{\alpha\beta}_{AA}(x,y) = - i g^2 \langle J^{\alpha}_A(x) J^{\beta}_{A}(y) \rangle\bigg\vert_{A=0}.
\eea

\section{Ward identities}
The consistent definition of the $\langle TJ_A J_A \rangle$ correlator requires the imposition of some Ward identities on it, that we are going to derive below. We start from the Ward identity to be satisfied by the axial vector current and then move to the conservation equation of the energy momentum tensor.

\subsection{Axial vector Ward identities}
 The axial vector Ward identity is given by
\bea
\partial^{x}_\alpha \, \Gamma_{AA}^{\mu\nu\alpha\beta}(z;x,y) = \partial^{x}_\alpha \left[ V_{55}^{\mu\nu\alpha\beta}(z;x,y) + W_{55}^{\mu\nu\alpha\beta}(z;x,y) \right].
\eea
The two terms in the previous equation take the form 
\bea
\partial^{x}_\alpha \, V_{55}^{\mu\nu\alpha\beta}(z;x,y) &=& (i \, g )^2 \, \partial^{x}_\alpha  \, \lag T_{f}^{\mu\nu} (z) J_A^{\alpha} (x) J_A^{\beta} (y) \rag \,, \label{WI1} \\
\partial^{x}_\alpha \, W_{55}^{\mu\nu\alpha\beta}(z;x,y) &=& g^{\alpha ( \mu} \Pi^{\nu)\beta}_{AA}(z,y)\,  \partial_\alpha^x \delta^{4}(x-z) + 2 m i\, \delta^4 (y-z) g^{\beta(\mu} \Pi_{AP}^{\nu )}(z, x) \nn \\
&-& g^{\mu\nu} \Pi^{\alpha\beta}_{AA}(x,y) \partial^{x}_\alpha \delta^4(x-z) - 2 m i\, g^{\mu\nu}[\delta^4(x-z) + \delta^4(y-z) ]\Pi_{AP}^{\beta}(x, y), \label{WI2}
\eea
while  $\Pi^{\alpha}_{AP}(x,y)$ is defined by
\bea
\Pi^{\alpha}_{AP}(x,y) = -i g^2 \langle J^{\alpha}_5(x) P(y) \rangle\bigg\vert_{A=0},
\eea
Here, $P$ denotes the pseudoscalar current $P \equiv \bar \psi \g_5 \psi$, and $\Pi^{\alpha}_{AP}\,,  \Pi^{\alpha\beta}_{AA}$ are related by the PCAC condition
\bea
2 i\,m \, \Pi^{\beta}_{AP}(x,y) = \partial_\alpha^x \, \Pi^{\alpha\beta}_{AA}(x,y) .
\label{PCAC}
\eea
The derivative of the correlator with the insertion of the free energy momentum tensor ($T_f$) can be calculated using functional techniques. For this purpose we consider the generating functional with the fermionic sources $\eta$ and $\bar \eta$ and the classical background sources $V^{\mu}$ and $A^{\mu}$ coupled respectively to the current operators $J_V ^{\mu}= \bar \psi \gamma^\mu \psi$ and $J_A^{\mu}= \bar \psi \gamma^\mu \gamma_5 \psi$
\bea
\langle T^{\mu\nu}_f(z) \rangle_{V,A,\eta,\bar\eta} = \int D \psi D \bar\psi \,\, T^{\mu\nu}_f(z) \, e^{i \mathcal{S}_k(\psi) + i \int d^4 x\, (g\, J_V \cdot V + g \,J_A \cdot A + \bar\psi \eta + \bar\eta \psi)}
\eea
and exploit the consequence of a chiral transformation on the corresponding Green's functions.\\
The functional integral must be invariant under a reparameterization of the integration variables, giving the identity
\bea
&&\int D \psi D \bar\psi \,\, T^{\mu\nu}_f(z) \, e^{i \,\mathcal{S}_k(\psi) + i \int d^4 x \, ( g\,J_V \cdot V + g\, J_A \cdot A + \bar\psi \, \eta + \bar\eta \, \psi)} =  \nn \\
&&\hspace{4cm} \int D \psi' D \bar\psi' \,\, T^{\mu\nu}_f(z)' \, e^{i \, \mathcal{S}_k(\psi') + i \int d^4  x \, (g\,J_V' \cdot V + g\,J_A' \cdot A + \bar\psi' \, \eta + \bar\eta \, \psi')}.
\label{FuncVar}
\eea
For a local  infinitesimal chiral transformation of the fermion fields defined by
\bea
\psi \rightarrow \psi' &=& \psi + i \, g \, \epsilon(x) \, \gamma_5 \, \psi , \\
\bar \psi \rightarrow \bar\psi' &=& \bar \psi + i \, g \, \epsilon(x) \, \bar \psi \, \gamma_5,
\eea
we can compute the variation of the action ${\mathcal S}$ and of $T^{\mu\nu}_p$ appearing on the the right hand side (r.h.s.) of eq.~(\ref{FuncVar}). The action changes as
\bea
\mathcal S_k(\psi')' = \mathcal S_k(\psi) + \int d^4 x \, \epsilon(x) (\partial_\alpha J^{\alpha}_A(x) - 2 i \, m P(x)),
\eea
whereas the vector and the axial-vector currents are obviously invariant
\bea
J_V^{\prime \mu} = J_V^{\mu}, \qquad J^{\prime \mu}_A  = J^{\mu}_A.
\eea
The variation of the free energy-momentum tensor is instead given by
\bea
\delta T^{\mu\nu}_f(z) = \frac{1}{2}\bigg[ J^{\mu}_A(z) \partial^{\nu}\epsilon(z) + J^{\nu}_A(z) \partial^{\mu}\epsilon(z)  \bigg] - g^{\mu\nu} \bigg[ J^{\lambda}_A(z) \partial_{\lambda} \epsilon(z) - 2 m i \epsilon(z) P(z)  \bigg].
\eea
We note that this change of variables is not a gauge transformation; $V$ and $A$ are therefore invariant. For this reason, using also the invariance of the two currents, the interaction terms $T_{i,A}$ and $T_{i,V}$ of the energy momentum tensor remain invariant as well. It follows that the variation of $T^{\mu\nu}_p(z)$ is due only to the free contribution shown above.\\
If we rewrite the infinitesimal parameter $\epsilon (z)$ as $\epsilon (z) = \int d^4 x \, \epsilon(x) \delta^4(z-x)$, the energy momentum variation can be recast in the following form
\bea
\delta T^{\mu\nu}_f (z) = \int d^4 x \, \epsilon(x) \, \mathcal H^{\mu\nu}(x,z),
\eea
where this definition of $\mathcal H^{\mu\nu}(x,z)$
\bea
&&\mathcal H^{\mu\nu}(x,z) =
\frac{1}{2}\, J^{\mu}_A(z) \, \partial^{\nu}_z \, \delta^4(z-x)
+ \frac{1}{2} \, J^{\nu}_A(z) \, \partial^{\mu}_z \, \delta^4(z-x) \nn \\
&& \hspace{3cm}- g^{\mu\nu} \left( J^{\lambda}_A(z) \, \partial^{z}_{\lambda} \, \delta^4(z-x)
- 2 i \, m P(z) \, \delta^4(z-x) \right)
\eea
will turn useful in the following. Given the chiral nature of the transformation, we include also the anomalous variation of the measure
\bea
D \psi' D \bar \psi' = D \psi D \bar\psi \, \exp \left\{i \int d^4 x \, \epsilon(x) a_n
\left[\frac{1}{3} \epsilon^{\alpha\beta\mu\nu}F_{\alpha\beta}^A F_{\mu\nu}^A +
\epsilon^{\alpha\beta\mu\nu}F_{\alpha\beta}^V F_{\mu\nu}^V
\right]
\right\}
\eea
where $a_n = \frac{g^2}{16 \pi^2}$ is the anomaly coefficient.
Expanding the r.h.s. of eq. (\ref{FuncVar}) to the first order in $\epsilon$ and taking into account the variation of the measure we obtain the Schwinger-Dyson equation
\bea
0 &=& \int d^4 x \, \epsilon (x)\,\int D\psi D\bar\psi \bigg\{i\,  T^{\mu\nu}_f(z) \bigg[ \partial_\alpha J^\alpha_A(x) - 2 m i P(x) + i g \bar\psi(x)\gamma_5 \eta(x)
+i g \bar\eta(x) \gamma_5 \psi(x)  \nn \\
&+&   a_n \left(\frac{1}{3} F^A(x) \tilde{F}^A(x) + F^V(x)\tilde{ F}^V(x)
\right) \bigg]+ \mathcal H^{\mu\nu}(x,z) \bigg\} e^{i \mathcal{S}_k(\psi) + i \int d^4  x \, ( g\,J_V \cdot V + g\, J_A \cdot A + \bar\psi \eta + \bar\eta \psi)}
\nn 
\label{axialWardId}
\eea
(with $F\tilde{F}\equiv  \epsilon^{\alpha\beta\mu\nu}F_{\alpha\beta} F_{\mu\nu}$).
The expression takes a simplified form if we set the sources $\eta, V$ and $\bar{\eta}$ to zero, and hence we obtain the anomalous Ward identity
\beq
i\langle T_f^{\mu\nu}(z)\partial\cdot J_A(x)\rangle_A = - 2  m \langle T^{\mu\nu}(z) P(x)\rangle_A
- i a_n \frac{1}{3}  F^A(x) \tilde{F}^A(x) \langle T_f^{\mu\nu}(z) \rangle_A - \langle{\mathcal H}^{\mu\nu}(x, z)\rangle_A.
\eeq

From Eq. (\ref{axialWardId}) we can extract Ward identities on correlation functions which contain one insertion of the energy-momentum tensor and several gauge currents just by
functional differentiation respect to the external sources. For example, taking a derivative of (\ref{axialWardId}) with respect to background field $A^{\mu}$  we obtain the constraint
\bea
&& \partial^{x}_\alpha \, \frac{ \delta} {\delta A^{\beta}(y)} \langle T^{\mu\nu}_f(z) J^{\alpha}_A(x) \rangle_{V,A,\eta,\bar\eta} \bigg\vert_{V,A,\eta,\bar\eta=0} =  \nn \\
&&\hspace{4cm}
\frac{\delta}{\delta A^{\beta}(y)} \bigg\{ 2 m i \langle T^{\mu\nu}_f(z) \, P(x) \rangle_{V,A,\eta,\bar\eta}
+  i \, \langle  \mathcal H^{\mu\nu}(x,z) \rangle_{V,A,\eta,\bar\eta} \bigg\} \bigg\vert_{V,A,\eta,\bar\eta=0}, \nonumber \\
\label{partialWI}
\eea
and performing explicitly the functional derivative we obtain the axial Ward identity
\beq
\partial^{x}_\alpha \, \langle T^{\mu\nu}_f(z) J^{\alpha}_A(x) J^{\beta}_A(y) \rangle  =  2 m i \langle T^{\mu\nu}_f(z) P(x) J^{\beta}_A(y) \rangle  +  i \langle  \mathcal H^{\mu\nu}(x,z) J^{\beta}_A(y) \rangle
\label{partialWI2}
\eeq
where the last term is given by
\bea
\langle  \mathcal H^{\mu\nu}(x,z) J^{\beta}_A(y) \rangle_{V,A,\eta,\bar\eta} \bigg\vert_{V,A,\eta,\bar\eta=0} &=&
(- i g^2)^{-1}\bigg\{\frac{1}{2} \Pi^{\mu\beta}_{AA}(z,y) \partial^{\nu}_z \delta^{4}(z-x) + \frac{1}{2} \Pi^{\nu\beta}_{AA}(z,y) \partial^{\mu}_z \delta^{4}(z-x) \nn \\
&-& g^{\mu\nu}\bigg[ \Pi^{\lambda\beta}_{AA}(z,y) \partial_{\lambda}^z \delta^{4}(z-x) - 2 m i\,\Pi^{\beta}_{AP}(z,y) \delta^{4}(z-x) \bigg] \bigg\}.
\label{HWI}
\eea
Notice that Eq. (\ref{partialWI2}) allows to derive indirectly the vacuum expectation value of the commutator of $T_f$ with $J_A$ by comparison with  the canonical expression
\bea
\partial^{x}_\alpha \, \langle T^{\mu\nu}_f(z) J^{\alpha}_A(x) J^{\beta}_A(y) \rangle  =  2 m i \langle T^{\mu\nu}_f(z) P(x) J^{\beta}_A(y) \rangle
+ \langle \big[ T^{\mu\nu}(z),J^\alpha_A(x)\big]g_{\alpha,0}\delta(x_0-z_0)J_A^\beta(y) \rangle
\label{partialWI3}
\eea
or
\beq
\langle \big[ T^{\mu\nu}(z),J^\alpha_A(x)\big]g_{\alpha,0}\delta(x_0-z_0)J_A^\beta (y) \rangle=
 i \langle  \mathcal H^{\mu\nu}(x,z) J^{\beta}_A(y) \rangle.
 \eeq
Proceeding with the functional differentiation one can derive further unrenormalized Ward identities for correlators of the form $T J_A J_A J_A$
\bea
(i g)^2 \, \partial^x_{\lambda} \, \langle T^{\mu\nu}_f(z)J_A^{\lambda}(x) J_A^{\alpha}(y)J_A^{\beta}(w)\rangle &=& (i g)^2 \, \langle T^{\mu\nu}_f(z) \, 2 m i P(x) \, J_A^{\alpha}(y)\, J_A^{\beta}(w)\rangle \nn \\
& & + \frac{8}{3} a_n \epsilon^{\alpha\beta\rho\sigma} \,
\partial_{\rho}\delta^4(x-y) \, \partial_{\sigma}\delta^4(x-w)\langle T^{\mu\nu}_f(z)\rangle \nonumber \\
&& + i (i g)^2 \, \langle \mathcal H^{\mu\nu}(x,z)J_A^{\alpha}(y)J_A^{\beta}(w)\rangle, \nonumber \\
\eea
which can be analyzed and checked in perturbation theory in a specific regularization scheme.
\subsection{The axial Ward identity in momentum space}
The Ward identity on the $\langle T J_A J_A \rangle$ vertex is extracted combining eqs.~({\ref{partialWI2}}) and (\ref{HWI}) with eqs.~(\ref{WI1}) and (\ref{WI2}) and it is explicitly given by
\bea
\partial^{x}_\alpha \, \Gamma_{AA}^{\mu\nu\alpha\beta}(z;x,y) &=& 2 m i (i \, g )^2 \, \lag T_{f}^{\mu\nu} (z) P(x) J_A^{\beta}(y) \rag + \bigg\{\frac{1}{2} \Pi^{\mu\beta}_{AA}(z,y) \partial^{\nu}_z \delta^{4}(z-x) \nn \\
&+& \frac{1}{2} \Pi^{\nu\beta}_{AA}(z,y) \partial^{\mu}_z \delta^{4}(z-x)
- g^{\mu\nu}\bigg[ \Pi^{\lambda\beta}_{AA}(z,y) \partial_{\lambda}^z \delta^{4}(z-x) - 2 m \,i \, \Pi^{\beta}_{AP}(z,y) \delta^{4}(z-x) \bigg] \bigg\} \nn \\
&+& g^{\alpha ( \mu} \Pi^{\nu)\beta}_{AA}(z,y)\,  \partial_\alpha^x \delta^{4}(x-z) + 2 m \,i\, \delta^4 (y-z) g^{\beta(\mu} \Pi_{AP}^{\nu )}(z, x) \nn \\
&-& g^{\mu\nu} \Pi^{\alpha\beta}_{AA}(x,y) \partial^{x}_\alpha \delta^4(x-z) - 2 m \,i\, g^{\mu\nu}[\delta^4(x-z) + \delta^4(y-z) ]\Pi_{AP}^{\beta}(x, y).
\eea
By defining
\beq
(2 \pi)^4\,  \delta^4(k-p-q) \, \Gamma_{AA}^{\mu\nu\alpha\beta}(k,p,q)=
\int d^4 x \, d^4 y \, d^4 z \, e^{-i \, k\cdot z + i \, p\cdot x + i \, q\cdot y}\,
\Gamma_{AA}^{\mu\nu\alpha\beta}(z;x,y)
\eeq
and
\beq
(2 \pi)^4 \, \delta^4(k-p-q) \, \Delta_{AP}^{\mu\nu\beta}(k,p,q)=
\int d^4 x \, d^4 y \, d^4z \, e^{-i \, k\cdot z + i \, p\cdot x + i \, q\cdot y}\, \lag T_{f}^{\mu\nu} (z) P(x) J_A^{\beta}(y) \rag,
\eeq
 we obtain its form in momentum space
\bea
-i p_\alpha \, \Gamma_{AA}^{\mu\nu\alpha\beta}(k,p,q) &=& 2 m i (i \, g )^2 \, \Delta_{AP}^{\mu\nu\beta}(k,p,q) + \bigg\{\frac{1}{2} i p^\nu\Pi^{\mu\beta}_{AA}(q) \nn \\
&+& \frac{1}{2}i p^\mu \Pi^{\nu\beta}_{AA}(q)
- g^{\mu\nu}\bigg[ i p_\lambda\Pi^{\lambda\beta}_{AA}(q) - 2 m \,i \, \Pi^{\beta}_{AP}(q)  \bigg] \bigg\} \nn \\
&-& i p_\alpha g^{\alpha ( \mu} \Pi^{\nu)\beta}_{AA}(q)\,  + 2 m \,i\,  g^{\beta(\mu} \Pi_{AP}^{\nu )}(p) \nn \\
&+& g^{\mu\nu} i p_\alpha\Pi^{\alpha\beta}_{AA}(q) - 2 m \,i\, g^{\mu\nu}\bigg[\Pi_{AP}^\beta(q) + \Pi_{AP}^{\beta}(p)\bigg].
\eea
We will be using this identity in the definition of the correlator with two axial-vector currents.
\subsection{Ward identity for the conservation of $T_{\mu\nu}$}
Moving to the the conservation equation of the energy momentum tensor, the derivation of the corresponding Ward identity involves the functional relation
(\ref{naiveWI}) which is given by
\bea
\frac{\partial}{\partial z^\nu}\Gamma_{AA}^{\mu\nu\alpha\beta} (z;x,y) &=&
- \frac{\partial}{\partial z_\mu}\, \delta^4(z-x)\, \Pi_{AA}^{\alpha\beta}(z,y) +
 g^{\alpha\mu} \, \frac{\partial}{\partial z^\lambda} \, \delta^4(z-x) \, \Pi_{AA}^{\lambda\beta}(z,y)\nonumber \\
&& - \frac{\partial}{\partial z_\mu}\, \delta^4(z-y) \, \Pi_{AA}^{\alpha\beta}(x,z) +
 g^{\beta\mu} \, \frac{\partial}{\partial z^\lambda} \, \delta^4 (z-y) \, \Pi_{AA}^{\lambda\alpha}(z,x),
\eea
which can be simplified using the PCAC relation (\ref{PCAC}). In momentum space it gives
\bea
k_{\nu}\Gamma_{AA}^{\mu\nu\alpha\beta}(p,q)=
(g^{\alpha\mu} \, k_{\nu} - g^\alpha_\nu \, p_{\mu}) \, \Pi_{AA}^{\beta\nu}(q)
+  (g^{\beta\mu}\, k_{\nu} - g^\beta_\nu \, q_{\mu})\, \Pi_{AA}^{\alpha\nu}(p).
\eea
The complete set of defining conditions of each vertex, beside the two Ward identities derived above, is the request of a symmetry on its $\mu,\nu$ indices, i.e.
$\Gamma_{AA}^{\mu\nu\alpha\beta}=\Gamma_{AA}^{\nu\mu\alpha\beta}$. We will be using these conditions in order to fix the entire structure of the correlator and check the consistency of a given regularization scheme.

\section{Diagrammatic expansion}
\begin{figure}[t]
\begin{center}
\includegraphics[scale=0.9]{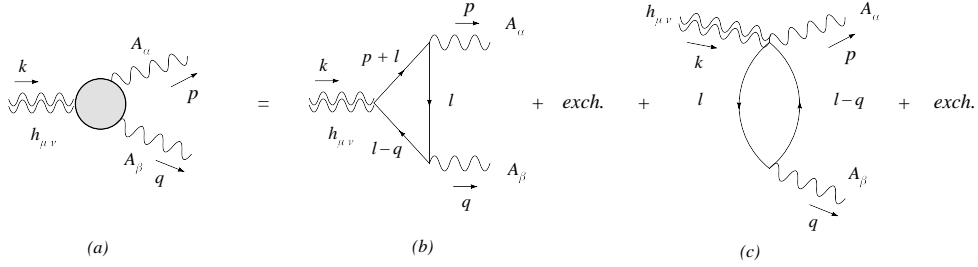}
\caption{\small The complete one-loop vertex (a) given by the sum of the 1PI contributions called $V_{55}
^{\mu \nu \a \b} (p,q)$ (b) and $W_{55}^{\mu \nu \a \b} (p,q)$ (c) with a graviton $h_{\mu\nu}$ in the initial state and two gauge bosons with axial-vector couplings $A_\a, A_\b$ in the final state. }
\label{axialvertex}
\end{center}
\end{figure}

The relevant diagrams responsible for the conformal anomaly are shown in Fig.~\ref{axialvertex} and take the form of eqs.~(\ref{V55c}) and (\ref{tbubble}). They consist of an amplitude with triangular topology (see Fig.~\ref{axialvertex}b) and of a bubble-like diagram (called  a ``t-bubble", see Fig.~\ref{axialvertex}c). This has the topology of a self-energy loop inserted on each of the gauge lines and attached from one side to the T vertex. These contributions are all of $O(g^2)$. At this point, we recall that the tree-level vertex with a graviton and a Dirac fermion, namely $V^{\prime \, \mu\nu}$, and the trilinear graviton-gauge boson-fermion coupling, i.e. $W_5^{\prime\,\mu\nu\alpha}$, induced by the two contributions $T_f$ and $T_{i A}$ are respectively given by
\bea
 V^{\prime \, \mu\nu}(k_1, k_2)&=&\frac{1}{4} \left[\gamma^\mu (k_1 + k_2)^\nu
+\gamma^\nu (k_1 + k_2)^\mu \right] - \frac{1}{2} g^{\mu \nu}
[\gamma^{\lambda}(k_1 + k_2)_{\lambda} - 2 m]
\label{Vprime}
\eea
\bea
W_5^{\prime\,\mu\nu\alpha} &=& -\frac{1}{2} (\gamma^\mu \gamma_5 g^{\nu\alpha}
+\gamma^\nu \gamma_5 g^{\mu\alpha}) +  g^{\mu \nu}\gamma^{\alpha} \gamma_5
\label{Wprime}
\eea
where $k_1$ and $k_2$ are generic momenta, incoming and outgoing, respectively.
Notice that the first contribution is vector-like, derived from (\ref{tfermionic}) and, naturally,  is the same appearing in the previous analysis of the $\langle TJJ \rangle$ correlator in \cite{Armillis:2009pq}.
The second one, $W_5^{\prime\,\mu\nu\alpha}$, due to (\ref{taxial}), differs from the analogous vertex $W^{\prime\,\mu\nu\alpha}$ appearing in the case of the $\langle TJJ \rangle$ correlator because of  the presence of the $\g_5$ matrix.\\
If we denote with $k$ the incoming momentum of  the graviton and with $p$ and $q$ the two outgoing momenta of the $A$ gauge bosons we obtain
\beq
(2 \, \pi)^4 \, \delta^4(k - p -q) \, V_{55}^{\mu\nu\alpha\beta}(p,q)\equiv \int d^4 x \, d^4 y \, d^4 z \, e^{- i k\cdot  z + i p\cdot x + i q\cdot y} \,
 \langle T_f^{\mu\nu}(z) \,  J_A^\a(x)  \,  J_A^\b(y)  \rangle
\eeq
\beq
(2 \, \pi)^4 \,  \delta^4(k - p -q)   \, W_{5\,5}^{\mu\nu\alpha\beta}(p,q)\equiv  \int d^4 x \, d^4 y \, d^4 z \, e^{- i k\cdot  z + i p\cdot x + i q\cdot y} \,
\langle T_{i A}^{\mu\nu}(z) J_A^\a(x) J_A^\b(y) \rangle.
\eeq
Explicitly
\bea
&&V_{5\,5}^{\mu\nu\alpha\beta}(p,q)=
-i g^2  \int \frac{d^4 l}{(2 \pi)^4}
\, \frac{{\rm tr}
\left\{V^{\prime\,\mu\nu}(l+p,l-q) (\lsl -\qsl +m)\gamma^{\beta}\gamma_5\,
(\lsl + m)\, \gamma^{\alpha}\gamma_5(\lsl +\psl + m)  \right\}}
{[l^2 - m^2] \, [(l-q)^2 - m^2] \, [(l+p)^2 - m^2] }\,,
\nn \\ \\
\label{V55}
&&W_{5\,5}^{\mu\nu\alpha\beta}(p,q)=
-i g^2  \int \frac{d^4 l}{(2 \pi)^4}
\, \frac{{\rm tr}
\left\{W_5^{\prime\,\mu\nu\alpha} \,
(\lsl + m)\, \gamma^{\beta}\gamma_5(\lsl +\qsl + m)  \right\}}
{[l^2 - m^2] \, [(l+q)^2 - m^2] },
\label{W55}
\eea
so that the complete one-loop amplitude (see Fig.~\ref{axialvertex}) is built up by symmetrizing on the external boson lines as
\bea
\Gamma_{AA}^{\mu\nu\alpha\beta}(p,q) = V_{5\,5}^{\mu\nu\alpha\beta}(p,q) + V_{5\,5}^{\mu\nu\beta\alpha}(q,p)+ W_{5\,5}^{\mu\nu\alpha\beta}(p,q) + W_{5\,5}^{\mu\nu\beta\alpha}(q,p).
\eea

\section{Tensor decomposition and naive manipulations}
As we have mentioned, the correlator is completely defined by a set of Ward identities, which amount to renormalization conditions which should be imposed in such a way 1) to respect its Bose symmetry and 2) the conservation of the fundamental currents of the theory. This is the case for all the anomalous correlators, both for chiral and conformal anomalies. At the same time, one needs a good regularization scheme in order to proceed with the actual implementation of these conditions, which could be obviously violated. This may require a (final) finite renormalization of the result in order to force the result to satisfy the original Ward identities. In this respect, various regularization schemes are available for chiral vertices, from a partially \cite{'tHooft:1972fi} to a totally anticommuting $\gamma_5$.
As we have already mentioned, in the computation of the correlator we have used DRED \cite{Siegel:1979wq}, with loop momenta computed in $D$ spacetime dimensions and traces performed in 4 dimensions, and we have verified all the Ward identities formally derived in this work. 

\subsection{Vanishing of the $TJ_VJ_A$ correlator}
We start our analysis by studying the  $TJ_VJ_A$  correlator.

For this reason we just recall that this specific correlation function can be extracted by the generating functional 
\beqa
\langle T_p^{\mu\nu}(z)\rangle_{V, A} &\equiv& \int D\psi D\bar{\psi} \,\,T^{\mu\nu}_p (z) \,\,e^{i \int d^4 x (\mathcal{L}_k(\psi) + g \,  J_V \cdot V(x) + g \, J_A\cdot A(x)) }\nonumber \\
&=& \langle T^{\mu\nu}_p \, e^{i \int d^4 x \, g \, (  \, J_V\cdot V(x) + J_A\cdot A(x))}\rangle.
\eeqa
Here we have introduced two independent sources $J_V$ and $J_A$. The corresponding correlators are obtained via functional variations respect to the background fields $V$ and $A$, namely
\beq
\Gamma_{VA}^{\mu\nu\alpha\beta} (z; x, y) \equiv \frac{ \delta^2 \lag T_p^{\mu\nu} (z) \rag_{V,A}}
{\delta V_{\alpha}(x)\delta A_{\beta}(y)} \bigg\vert_{V,A=0}
= V_{5}^{\mu\nu\alpha\beta} (z; x, y) + W_{5}^{\mu\nu\alpha\beta} (z; x, y).
\label{VAcorrelator}
\eeq
whose expressions in momentum space are (for the direct and the exchange contributions)
\bea &&V_{5\,dir}^{\mu\nu\alpha\beta}(p,q)= - (-ig)^2 i^3  \int
\frac{d^4 l}{(2 \pi)^4} \, \frac{{\rm tr}
\left\{V^{\prime\,\mu\nu}(l+p,l-q) (\lsl -\qsl +m)\gamma^{\beta}\,
(\lsl + m)\, \gamma^{\alpha}\gamma_5(\lsl +\psl + m)  \right\}}
{[l^2 - m^2] \, [(l-q)^2 - m^2] \, [(l+p)^2 - m^2] }\,,
\label{V5dir}
\nn \\ \\
&&V_{5\,ex}^{\mu\nu\alpha\beta}(p,q)= - (-ig)^2 i^3  \int
\frac{d^4 l}{(2 \pi)^4} \, \frac{{\rm tr}
\left\{V^{\prime\,\mu\nu}(l-p,l+q) (\lsl -\psl
+m)\gamma^{\alpha}\gamma_5\, (\lsl + m)\, \gamma^{\beta}(\lsl
+\qsl + m)  \right\}} {[l^2 - m^2] \, [(l+q)^2 - m^2] \, [(l-p)^2
- m^2] }\,,
\label{V5ex}
\nn \\ \\
&&W_{5\,dir}^{\mu\nu\alpha\beta}(p,q)= - (-ig)^2 i^3  \int
\frac{d^4 l}{(2 \pi)^4} \, \frac{{\rm tr}
\left\{W^{\prime\,\mu\nu\alpha}\gamma_5 (\lsl +m)\gamma^{\beta}\, (\lsl + \qsl +m)\, \right\}} {[l^2 - m^2] \, [(l+q)^2 - m^2] }\,,
 \label{W5dir}\\
&&W_{5\,ex}^{\mu\nu\alpha\beta}(p,q)= - (-ig)^2 i^3  \int
\frac{d^4 l}{(2 \pi)^4} \, \frac{{\rm tr}
\left\{W^{\prime\,\mu\nu\beta} (\lsl +m)\gamma^{\alpha}\gamma_5\, (\lsl + \psl +m)\, \right\}} {[l^2 - m^2] \, [(l+p)^2 - m^2] }\,,
\label{W5ex}
\eea
and where the vertices $V^{\prime\,\mu\nu}$ and  $W^{\prime\,\mu\nu\alpha}$ are defined as
\bea
 V^{\prime \, \mu\nu}(k_1, k_2)&=&\frac{1}{4} \left[\gamma^\mu (k_1 + k_2)^\nu
+\gamma^\nu (k_1 + k_2)^\mu \right] - \frac{1}{2} g^{\mu \nu}
[\gamma^{\lambda}(k_1 + k_2)_{\lambda} - 2 m],\label{Vprime2}\\
W^{\prime\,\mu\nu\alpha} &=& -\frac{1}{2} (\gamma^\mu  g^{\nu\alpha}
+\gamma^\nu  g^{\mu\alpha}) +  g^{\mu \nu}\gamma^{\alpha}.\label{Wprime2}
\eea
We will use the same trick used for the proof of Furry's theorem to show the vanishing of this correlator, which is formally divergent and therefore ill-defined. For this reason one needs some external Ward identities in order to resolve its structure. For the $TJ_V J_A$ vertex the situation is quite peculiar since one can show, using DRED and by allowing momentum shifts, that
the three Ward identities are indeed homogeneous
  \beq
  k_\mu\Gamma_{VA}^{\mu\nu\alpha\beta}=
  p_\alpha\Gamma_{VA}^{\mu\nu\alpha\beta}=
  q_\beta\Gamma_{VA}^{\mu\nu\alpha\beta}=0,
\eeq
while the properties of symmetry of the correlator are respected. Obviously, this indicates that there is a regularization scheme in which the anomaly of the axial-vector current $J_A$ does not appear. 
A closer inspection shows that this result is caused by a cancellation between the direct and the exchange contribution, since the
$\epsilon$-tensor is present in each of the two (direct and exchange) diagrams contributing to the vertex, but not in their sum. Indeed, this clearly seems to indicate that this correlator may be vanishing identically. A second argument, based on charge conjugation invariance brings to identical conclusions.

For this reason, we take the expression of the triangle diagram and insert the identity $C^{-1}\,C = 1$ - involving the charge conjugation matrix $C$ between every $\gamma$ matrix - together with the relations
\bea
C\,\gamma^{\mu}\,C^{-1}= -(\gamma^{\mu})^T, \qquad\qquad
C\,\gamma_5\,C^{-1}= \gamma_5,
\eea
so that the trace in eq.~(\ref{V5dir}) becomes
\bea
\mathcal T &=& {\rm tr} \left\{\tilde V^{\prime\,\mu\nu}(l+p,l-q)^T (\lsl -\qsl -m)^T(\gamma^{\beta})^T\,(\lsl - m)^T\, (\gamma^{\alpha})^T\gamma_5(\lsl +\psl - m)^T  \right\} \nn \\
&=& - {\rm tr} \left\{\tilde V^{\prime\,\mu\nu}(l+p,l-q) (\lsl
+\psl -m)\gamma^{\alpha}\gamma_5\,(\lsl - m)\,
\gamma^{\beta}\,(\lsl -\qsl - m)  \right\} \label{tracegamma}
\eea
where $\tilde V^{\prime\,\mu\nu}$ differs from $V^{\prime\,\mu\nu}$ only for the sign of the mass term
\bea
\tilde V^{\prime \, \mu\nu}(k_1, k_2)=\frac{1}{4} \left[\gamma^\mu (k_1 +
k_2)^\nu +\gamma^\nu (k_1 + k_2)^\mu \right] - \frac{1}{2} g^{\mu
\nu} [\gamma^{\lambda}(k_1 + k_2)_{\lambda} + 2 m].
\eea
Changing the integration variable $l\rightarrow -l$ in eq.(\ref{tracegamma}) we get
\bea
\mathcal T = - {\rm tr} \left\{ V^{\prime\,\mu\nu}(l-p,l+q) (\lsl
-\psl +m)\gamma^{\alpha}\gamma_5\,(\lsl + m)\,
\gamma^{\beta}\,(\lsl +\qsl + m)  \right\}, \label{piece1}
\eea
while the three denominators in eq.(\ref{V5dir}) change according to
\bea
\frac{1}{[l^2 - m^2] \, [(l-q)^2 - m^2] \, [(l+p)^2 - m^2] }
\rightarrow \frac{1}{[l^2 - m^2] \, [(l+q)^2 - m^2] \, [(l-p)^2 -
m^2] }. \label{piece2}
\eea
Combining eq.(\ref{piece1}) and (\ref{piece2}) it is easy to recognize that
\bea
V_{5\,dir}^{\mu\nu\alpha\beta}(p,q) = -
V_{5\,ex}^{\mu\nu\alpha\beta}(p,q)
\eea
so that the sum of the two triangles vanishes.\\
The last point to check in order to be sure of the vanishing of the vertex concerns the contributions from the t-bubble diagrams. These have been defined in eq.(\ref{W5dir}) and (\ref{W5ex}) and their topology is the one showed in Fig.~\ref{axialvertex}c. These are both separately equal to zero because they consists of a combination of 2-point functions of the form $\Pi^{\alpha\beta}_{VA}(p)$ given by
\bea
\Pi^{\alpha\beta}_{VA}(p) = - g^2 \, \int
\frac{d^4 l}{(2 \pi)^4} \, \frac{{\rm tr}
\left\{ \gamma^{\alpha}\gamma_5\, (\lsl +m)\gamma^{\beta}\, (\lsl + \psl +m)\, \right\}} {[l^2 - m^2] \, [(l+p)^2 - m^2] }
\eea
which are also identically vanishing.

\begin{table}
$$
\begin{array}
{|c 
 | 
 c 
 |
c 
 |
c 
|
c
|
c 
|}
\hline
& & & & & \\[-.5cm]
\begin{array}[t]{c}
p^{\mu} p^{\nu} p^{\alpha} p^{\beta}\\
q^{\mu} q^{\nu} q^{\alpha} q^{\beta}
\end{array}
&
\begin{array}[t]{c}
p^{\mu} p^{\nu} p^{\alpha} q^{\beta}\\
p^{\mu} p^{\nu} q^{\alpha} p^{\beta}\\
p^{\mu} q^{\nu} p^{\alpha} p^{\beta}\\
q^{\mu} p^{\nu} p^{\alpha} p^{\beta}
\end{array}
&
\begin{array}[t]{c}
p^{\mu} p^{\nu} q^{\alpha} q^{\beta}\\
p^{\mu} q^{\nu} p^{\alpha} q^{\beta}\\
q^{\mu} p^{\nu} p^{\alpha} q^{\beta}
\end{array}
&
\begin{array}[t]{c}
p^{\mu} q^{\nu} q^{\alpha} p^{\beta}\\
q^{\mu} p^{\nu} q^{\alpha} p^{\beta}\\
q^{\mu} q^{\nu} p^{\alpha} p^{\beta}
\end{array}
&
\begin{array}[t]{c}
p^{\mu} q^{\nu} q^{\alpha} q^{\beta}\\
q^{\mu} p^{\nu} q^{\alpha} q^{\beta}\\
q^{\mu} q^{\nu} p^{\alpha} q^{\beta}\\
q^{\mu} q^{\nu} q^{\alpha} p^{\beta}
\end{array}
&
\begin{array}[t]{c}
g^{\mu\nu}g^{\alpha\beta}\\
g^{\alpha\mu}g^{\beta\nu}\\
g^{\alpha\nu}g^{\beta\mu}
\end{array}
\\[2.1cm]
\hline
& & & & & \\[-.5cm]
\begin{array}{c}
p^{\mu} p^{\nu} g^{\alpha\beta}\\
p^{\mu} q^{\nu} g^{\alpha\beta}\\
q^{\mu} p^{\nu} g^{\alpha\beta}\\
q^{\mu} q^{\nu} g^{\alpha\beta}
\end{array}
&
\begin{array}{c}
p^{\beta} p^{\nu} g^{\alpha\mu}\\
p^{\beta} q^{\nu} g^{\alpha\mu}\\
q^{\beta} p^{\nu} g^{\alpha\mu}\\
q^{\beta} q^{\nu} g^{\alpha\mu}
\end{array}
&
\begin{array}{c}
p^{\beta} p^{\mu} g^{\alpha\nu}\\
p^{\beta} q^{\mu} g^{\alpha\nu}\\
q^{\beta} p^{\mu} g^{\alpha\nu}\\
q^{\beta} q^{\mu} g^{\alpha\nu}
\end{array}
&
\begin{array}{c}
p^{\alpha} p^{\nu} g^{\beta\mu}\\
p^{\alpha} q^{\nu} g^{\beta\mu}\\
q^{\alpha} p^{\nu} g^{\beta\mu}\\
q^{\alpha} q^{\nu} g^{\beta\mu}
\end{array}
&
\begin{array}{c}
p^{\mu} p^{\alpha} g^{\beta\nu}\\
p^{\mu} q^{\alpha} g^{\beta\nu}\\
q^{\mu} p^{\alpha} g^{\beta\nu}\\
q^{\mu} q^{\alpha} g^{\beta\nu}
\end{array}
&
\begin{array}{c}
p^{\alpha} p^{\beta} g^{\mu\nu}\\
p^{\alpha} q^{\beta} g^{\mu\nu}\\
q^{\alpha} p^{\beta} g^{\mu\nu}\\
q^{\alpha} q^{\beta} g^{\mu\nu}
\end{array}
\\ [1.2cm]\hline
\label{WI}
\end{array}
$$
\caption{\small The 43 tensor monomials called $l_i^{\mu\nu\alpha\beta}(p,q)$ built up from the metric tensor and the two independent momenta $p$ and $q$ into which a general fourth rank tensor can be expanded.
\label{monomials}}
\end{table}

\subsection{The computation of the $\langle TJ_A J_A \rangle $ correlator}
 We now going to address the computation of the $TJ_A J_A$ vertex, but prior to that we briefly review the vector/vector case. 
 As discussed in \cite{Giannotti:2008cv} and in \cite{Armillis:2009pq} the full one-loop amplitude with the energy momentum tensor coupled to two vector currents, $\Gamma^{\mu\nu\alpha\beta}_{VV}$, can be expanded on the basis provided by the 43 monomial tensors $ l_i^{\mu \nu \a \b} (p,q)$ listed in Tab.~\ref{monomials}
\bea
\Gamma_{VV}^{\mu\nu\alpha\beta} (p,q) = \, \sum_{i=1}^{43} \, A_i(k^2,p^2,q^2) \, l_i^{\mu \nu \a \b} (p,q),
\eea
whose form factors $A_i (k^2,p^2,q^2)$ are not all convergent, since the amplitude has total mass dimension equal to $2$. It has been shown in \cite{Armillis:2009pq} that they can be divided into $3$ groups:

\begin{itemize}
\item[a)]  $A_1 \leq A_i \leq A_{16}$ - multiplied by a product of four momenta, they have mass dimension $-2$ and therefore are UV finite;
\item [b)] $A_{17} \leq A_i \leq A_{19}$ - these have mass dimension $2$ since the four Lorentz indices of the amplitude are carried by two metric tensors
\item [c)] $A_{20} \leq A_i \leq A_{43}$ - they appear next to a metric tensor and two momenta, have mass dimension $0$ and are divergent.
\end{itemize}
In \cite{Giannotti:2008cv}  the $43$ invariant amplitudes $A_i (k^2,p^2,q^2)$ have been cleverly reduced to the $13$
named $F_i (k^2,p^2,q^2)$. A similar result is obtained in \cite{Armillis:2009pq} using a different intermediate basis. This reorganization of the amplitude shows conclusively that the effective action of theories with conformal anomalies
is affected  by anomaly poles which contain the entire signature of the anomaly \cite{Armillis:2009im}.

As we are going to show, a similar result holds also for the $\langle TJ_A J_A \rangle$ vertex. At the same time, we are going to demonstrate the appearance only of conformal anomalies, since the mixed anomalies cancel, and present the complete expression of this vertex.
\begin{table}
$$
\begin{array}{|c|c|}\hline
i & t_i^{\mu\nu\alpha\beta}(p,q)\\ \hline\hline
1 &
\left(k^2 g^{\mu\nu} - k^{\mu } k^{\nu}\right) u^{\alpha\beta}(p.q)\\ \hline
2 &
\left(k^2g^{\mu\nu} - k^{\mu} k^{\nu}\right) w^{\alpha\beta}(p.q)  \\ \hline
3 & \left(p^2 g^{\mu\nu} - 4 p^{\mu}  p^{\nu}\right)
u^{\alpha\beta}(p.q)\\ \hline
4 & \left(p^2 g^{\mu\nu} - 4 p^{\mu} p^{\nu}\right)
w^{\alpha\beta}(p.q)\\ \hline
5 & \left(q^2 g^{\mu\nu} - 4 q^{\mu} q^{\nu}\right)
u^{\alpha\beta}(p.q)\\ \hline
6 & \left(q^2 g^{\mu\nu} - 4 q^{\mu} q^{\nu}\right)
w^{\alpha\beta}(p.q) \\ \hline
7 & \left[p\cdot q\, g^{\mu\nu}
-2 (q^{\mu} p^{\nu} + p^{\mu} q^{\nu})\right] u^{\alpha\beta}(p.q) \\ \hline
8 & \left[p\cdot q\, g^{\mu\nu}
-2 (q^{\mu} p^{\nu} + p^{\mu} q^{\nu})\right] w^{\alpha\beta}(p.q)\\ \hline
9 & \left(p\cdot q \,p^{\alpha}  - p^2 q^{\alpha}\right)
\big[p^{\beta} \left(q^{\mu} p^{\nu} + p^{\mu} q^{\nu} \right) - p\cdot q\,
(g^{\beta\nu} p^{\mu} + g^{\beta\mu} p^{\nu})\big]  \\ \hline
10 & \big(p\cdot q \,q^{\beta} - q^2 p^{\beta}\big)\,
\big[q^{\alpha} \left(q^{\mu} p^{\nu} + p^{\mu} q^{\nu} \right) - p\cdot q\,
(g^{\alpha\nu} q^{\mu} + g^{\alpha\mu} q^{\nu})\big]  \\ \hline
11 & \left(p\cdot q \,p^{\alpha} - p^2 q^{\alpha}\right)
\big[2\, q^{\beta} q^{\mu} q^{\nu} - q^2 (g^{\beta\nu} q^ {\mu}
+ g^{\beta\mu} q^{\nu})\big]  \\ \hline
12 & \big(p\cdot q \,q^{\beta} - q^2 p^{\beta}\big)\,
\big[2 \, p^{\alpha} p^{\mu} p^{\nu} - p^2 (g^{\alpha\nu} p^ {\mu}
+ g^{\alpha\mu} p^{\nu})\big] \\ \hline
13 & \big(p^{\mu} q^{\nu} + p^{\nu} q^{\mu}\big)g^{\alpha\beta}
+ p\cdot q\, \big(g^{\alpha\nu} g^{\beta\mu}
+ g^{\alpha\mu} g^{\beta\nu}\big) - g^{\mu\nu} u^{\alpha\beta} \\
& -\big(g^{\beta\nu} p^{\mu}
+ g^{\beta\mu} p^{\nu}\big)q^{\alpha}
- \big (g^{\alpha\nu} q^{\mu}
+ g^{\alpha\mu} q^{\nu }\big)p^{\beta}  \\ \hline
\end{array}
$$
\caption{The 13 fourth rank tensors $t_i^{\mu\nu\alpha\beta}(p,q)$ satisfying the vector current conservation on the external lines with momenta $p$ and $q$. \label{genbasis13}}
\end{table}

To illustrate this point, we observe that the insertion of the non-chiral component of $T^{\mu\nu}$ (represented by $T_f^{\mu\nu}$)  in the correlator $V_{55}$, defines one of the two subamplitudes which may potentially generate mixed anomalies. On the other hand, it is however obvious - by a glance at the structure of the correlator - that we could remove symmetrically the chiral matrix all together. Therefore, the $\langle TJ_A J_A \rangle$ correlator can be split in two terms, the first being the correlator with two vector currents called $TJ_VJ_V$, while the second  is an extra contribution, proportional to the fermion mass $m$, denoted by $\Omega$
\bea
\Gamma_{AA}^{\mu\nu\alpha\beta}(p,q) = \Gamma_{VV}^{\mu\nu\alpha\beta}(p,q) + \Omega^{\mu\nu\alpha\beta}(p,q). \label{GammaSplit}
\eea
The explicit computation of the correlator with two vector currents $\Gamma_{VV}^{\mu\nu\alpha\beta}$ can be borrowed from \cite{Armillis:2009pq}, but the computation of the extra terms is very involved, due to the need to select a specific number of tensor structures in its expansion. Notice that the decomposition in eq.~(\ref{GammaSplit}) is particularly useful because shows that the vector and axial-vector cases coincide in the chiral limit, i.e. for $\Omega^{\mu\nu\alpha\beta}=0$. \\
As we have just mentioned above, the amplitude $\Gamma^{\mu\nu\alpha\beta}_{VV}$ can be expanded in the reduced basis given in Tab. \ref{genbasis13} 
\bea
\Gamma_{VV}^{\mu\nu\alpha\beta}(p,q) =  \, \sum_{i=1}^{13} F_i (s; s_1, s_2,m^2)\ t_i^{\mu\nu\alpha\beta}(p,q)\,,
\label{Gamt}
\eea
where the invariant amplitudes $F_i (s; s_1, s_2,m^2)$ are functions of the kinematical invariants $s=k^2=(p+q)^2$, $s_1=p^2$, $s_2=q^2$. Their explicit expressions in the general case have been given in \cite{Armillis:2009pq}. In the simplest case, i.e. for an internal zero mass fermion ($m=0$) and on-shell photons on the external lines ($s_1=s_2=0$),  the only non-vanishing  $F_i(s; s_1, s_2,m^2)$ are given by
\bea
F_{1} (s, 0, 0, 0) &=& - \frac{g^2}{18 \pi^2  s}, \\
F_{3} (s, 0, 0, 0) &=&  F_{5} (s, 0, 0, 0) = - \frac{g^2}{144 \pi^2 \, s}, \\
F_{7} (s, 0, 0, 0) &=& -4 \, F_{3} (s, 0, 0, 0), \\
F_{13, R} (s, 0, 0, 0) &=& - \frac{g^2}{144 \pi^2} \, \left[ 12 \log \left(-\frac{s}{\mu^2}\right) - 35\right],
\label{formfts}
\eea
(with $s <0$) where $F_{13}$ is affected by charge renormalization (with a scale $\mu$). As we are going to discuss next, $F_1$ is the only form factor contributing to the trace anomaly in the massless case, and contains an anomaly pole. In this sense we can say that the pole {\em saturates} the anomaly and completely accounts for it.
In \cite{Giannotti:2008cv} this $1/s$ terms is identified by a spectral analysis of the correlator, while the same structure emerges form the complete expressions of the form factors derived in  \cite{Armillis:2009pq} and presented above. 

Coming instead to the new contribution  $\Omega^{\mu\nu\alpha\beta}$ appearing in eq.~(\ref{GammaSplit}), this can be written as
\bea
\Omega^{\mu\nu\alpha\beta}(p,q) = \Omega_{V}^{\mu\nu\alpha\beta}(p,q) + \Omega_{V}^{\mu\nu\beta\alpha}(q,p)+ \Omega_{W}^{\mu\nu\alpha\beta}(p,q) + \Omega_{W}^{\mu\nu\beta\alpha}(q,p),
\label{omegaall}
\eea
where the  amplitudes $\Omega_{V}^{\mu\nu\alpha\beta}$ and $\Omega_{W}^{\mu\nu\alpha\beta}$ are given by
\bea
&&\Omega_{V}^{\mu\nu\alpha\beta}(p,q)=
-2 m (-i g^2)  \int \frac{d^4 l}{(2 \pi)^4}
\, \frac{{\rm tr}
\left\{V^{\prime\,\mu\nu}(l+p,l-q) (\lsl -\qsl +m)\gamma^{\beta} \gamma^{\alpha}(\lsl +\psl + m) \right\}}
{[l^2 - m^2] \, [(l-q)^2 - m^2] \, [(l+p)^2 - m^2] }\,,
\label{OmegaV}
\nn \\ \\
&&\Omega_{W}^{\mu\nu\alpha\beta}(p,q)=
-2m(-i g^2)  \int \frac{d^4 l}{(2 \pi)^4}
\, \frac{{\rm tr}
\left\{W^{\prime\,\mu\nu\alpha} \,\gamma^{\beta}(\lsl +\qsl + m)  \right\}}
{[l^2 - m^2] \, [(l+q)^2 - m^2] },
\label{OmegaW}
\eea
with the $V^{\prime\,\mu\nu}$ and $W^{\prime\,\mu\nu\alpha}$ defined in eqs~(\ref{Vprime2}) and (\ref{Wprime2}). The remaining two terms in eq.~(\ref{omegaall}) are simply the Bose symmetric amplitudes obtained exchanging the indices $\alpha$ and $\beta$ and the momenta $p$ and $q$ of (\ref{OmegaV}) and (\ref{OmegaW}).
The extra term $\Omega^{\mu\nu\alpha\beta}$ can be expanded on the basis provided by the 43 monomial tensors $l_i^{\mu \nu \a \b} (p,q)$ listed in Tab.~\ref{monomials}
\bea
\Omega^{\mu\nu\alpha\beta}(p,q) = \, \sum_{i=1}^{43} \, E_i(k^2,p^2,q^2,m^2) \, l_i^{\mu \nu \a \b} (p,q),
\eea
where the form factors $E_i (k^2,p^2,q^2,m^2)$ are some functions of the kinematical variables and of the mass of the fermion in the loop. This needs to be identified by a direct inspection.
The explicit computation shows that not all the 43 invariant amplitudes $E_i (k^2,p^2,q^2,m^2)$ are really present in this expansion and therefore the surviving ones can be appropriately combined in a lower number of composite tensor structures. This result can be organized in a more compact form after introducing a new tensor basis whose elements $f_i^{\mu\nu\alpha\beta}(p,q)$ ($i=1,\dots, 9$) are listed in Tab.\ref{genbasis}. We obtain
\bea
\Omega^{\mu\nu\alpha\beta}(p,q) = \, \sum_{i=1}^{9} \, R_i (s,s_1,s_2,m^2) \, f_i^{\mu \nu \a \b} (p,q),
\label{OmegaR}
\eea
where the invariant amplitudes $R_i (s,s_1,s_2,m^2)$ depend on the kinematical variables $s = k^2 = (p+q)^2$, $s_1 = p^2$, $s_2 = q^2$ besides the fermion mass $m$.\\
Three of the nine tensors are Bose symmetric, namely,
\bea
f_i^{\mu\nu\alpha\beta}(p,q) = f_i^{\mu\nu\beta\alpha}(q,p) \,,\qquad i=1,6,9\,,
\eea
while the remaining ones form three pairs related by Bose symmetry
\bea
&&f_2^{\mu\nu\alpha\beta}(p,q) = f_3^{\mu\nu\beta\alpha}(q,p)\,,\\
&&f_4^{\mu\nu\alpha\beta}(p,q) = f_5^{\mu\nu\beta\alpha}(q,p)\,,\\
&&f_7^{\mu\nu\alpha\beta}(p,q) = f_8^{\mu\nu\beta\alpha}(q,p)\,.
\eea
This basis is particularly useful because only the first three of the nine tensors have a non-zero trace
\bea
g_{\mu \nu} f_1^{\mu \nu \alpha \beta}(p,q) &=& 3 k^2 g^{\alpha\beta}\,,\\
g_{\mu \nu} f_2^{\mu \nu \alpha \beta}(p,q) &=& g_{\mu \nu} f_3^{\mu \nu \alpha \beta}(p,q) = 2(p^{\alpha}q^{\beta}- p^{\beta}q^{\alpha}) \,,
\eea

while the remaining six tensors are traceless
\bea
g_{\mu\nu}f_i^{\mu\nu\alpha\beta}(p,q) = 0 \,,\qquad i=4,5,6,7,8,9 \,.
\eea
At this point, the goal is to express the amplitude $\Omega^{\mu\nu\alpha\beta} (p,q)$ in an analytical form. We start from the evaluation of the integrals in eqs.~(\ref{OmegaV}) and (\ref{OmegaW}), obtaining the form factors $E_i$. At a second stage we map them into the new parameterization defined in eq.~(\ref{OmegaR}),  determining in this way the coefficients $R_i$.
The relations between the two sets $\{E_i\}_{i=1,\dots,43}$ and $\{R_i\}_{i=1,\dots,9}$, for the most general external momenta are 
\bea
R_1 &=& \frac{1}{3 k^2}\left(E_{20}\,p^2 + 2 E_{21}\,p\cdot q + E_{23}\,q^2 + 4 E_{17} + 2 E_{18}\right), \\
R_2 &=& E_{26}, \\
R_3 &=& E_{33}, \\
R_4 &=& E_{26}, \\
R_5 &=& E_{33}, \\
R_6 &=& \frac{E_{18}}{p\cdot q}, \\
R_7 &=& -\frac{1}{12 k^2}\left(E_{20}\,p^2 + 2 E_{21}\,p\cdot q + E_{23}\,q^2 + 4 E_{17} + 2 E_{18}\right) -\frac{E_{20}}{4}, \\
R_8 &=& -\frac{1}{12 k^2}\left(E_{20}\,p^2 + 2 E_{21}\,p\cdot q + E_{23}\,q^2 + 4 E_{17} + 2 E_{18}\right) -\frac{E_{23}}{4}, \\
R_9 &=& -\frac{1}{6 k^2}\left(E_{20}\,p^2 + 2 E_{21}\,p\cdot q + E_{23}\,q^2 + 4 E_{17} + 2 E_{18}\right) +\frac{E_{18}}{2 p\cdot q} -\frac{E_{21}}{2},
\eea
\begin{table}
$$
\begin{array}{|c|c|}\hline
i & f_i^{\mu\nu\alpha\beta}(p,q)\\ \hline\hline
1 &
\left(k^2 g^{\mu\nu} - k^{\mu } k^{\nu}\right) g^{\alpha\beta}\\ \hline
2 &
p^{\nu}q^{\beta}g^{\alpha\mu} + p^{\mu}q^{\beta}g^{\alpha\nu} - p^{\nu}q^{\alpha}g^{\beta\mu} - p^{\mu}q^{\alpha}g^{\beta\nu}  \\ \hline
3 &
p^{\alpha}q^{\nu}g^{\beta\mu} + p^{\alpha}q^{\mu}g^{\beta\nu} - p^{\beta}q^{\nu}g^{\alpha\mu} - p^{\beta}q^{\mu}g^{\alpha\nu}
\\ \hline
4 &
p^{\nu}p^{\beta}g^{\alpha\mu} + p^{\mu}p^{\beta}g^{\alpha\nu} - p^{\nu}p^{\alpha}g^{\beta\mu} - p^{\mu}p^{\alpha}g^{\beta\nu}
\\ \hline
5 &
q^{\alpha}q^{\nu}g^{\beta\mu} + q^{\alpha}q^{\mu}g^{\beta\nu} - q^{\beta}q^{\nu}g^{\alpha\mu} - q^{\beta}q^{\mu}g^{\alpha\nu}
\\ \hline
6 &
(p^{\mu}q^{\nu}+q^{\mu}p^{\nu})g^{\alpha\beta}+ p\cdot q\,(g^{\alpha\nu}g^{\beta\mu}+g^{\alpha\mu}g^{\beta\nu}-g^{\alpha\beta}g^{\mu\nu})
\\ \hline
7 &
\left(p^2 g^{\mu\nu} - 4 p^{\mu } p^{\nu}\right) g^{\alpha\beta}
\\ \hline
8 &
\left(q^2 g^{\mu\nu} - 4 q^{\mu } q^{\nu}\right) g^{\alpha\beta}
\\ \hline
9 &
\left(p\cdot q g^{\mu\nu} - 2 (q^{\mu } p^{\nu} + p^{\mu } q^{\nu} )\right) g^{\alpha\beta}
\\ \hline
\end{array}
$$
\caption{Basis of 9 fourth rank tensors called $f_i^{\mu\nu\alpha\beta}(p,q)$. \label{genbasis}}
\label{reduced}
\end{table}
where all the dependence on the kinematical invariants $k^2,p^2,q^2$ and $m^2$ appearing in the sets $R_i$ and $E_i$ has been omitted. The explicit expressions in DRED of the form factors $R_i$ have been collected in Appendix \ref{RFormFactors} and represent an important step in the computation of the $\langle TJ_AJ_A \rangle $ correlator. These form factors are affected by the usual ultraviolet singularities, which
in a renormalizable theory would be removed by standard renormalization counterterms. In our case they turn out to be proportional to 2-point functions. 

Except for these possible counterterms, the main techniques and methods used in this analysis remain invariant and are of an easy application also in the case of the Standard Model. Notice, in particular, that the main equation (\ref{GammaSplit}) implies that the non-renormalizable contributions are proportional to mass corrections contributing to $\Omega$, and the non-renormalizable terms indeed involve correlators of two axial-vector currents, as just mentioned above. The renormalization of the first contribution $\Gamma_{VV}$ is canonical, and is attributed to the form factor $F_{13}$ of Eq. (\ref{formfts}), which is induced by a renormalization of 2-point functions of vector currents. 
\begin{figure}[t]
\begin{center}
\includegraphics[scale=0.9]{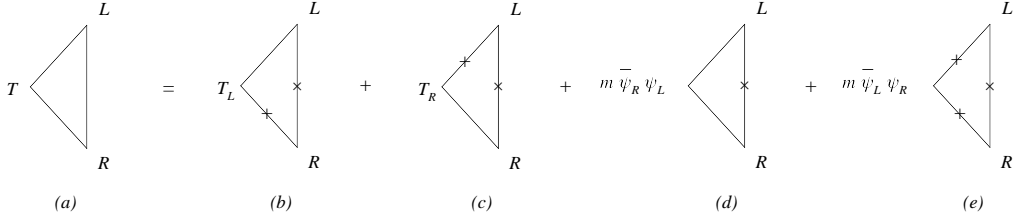}
\caption{\small Chiral decomposition of the correlator. }
\label{TLR}
\end{center}
\end{figure}


Before coming to the analysis of the other vertices, in closing this section we just remark that our analysis in the V/A basis can be rewritten completely in terms of chiral L/R currents, since the following relations hold for nonzero $m$
\bea
\langle  T J_V J_V \rangle &=& \langle  T J_L J_L \rangle + \langle  T J_R
J_R \rangle + \langle  T J_L J_R \rangle + \langle  T J_R J_L \rangle, \\
\langle  T J_A J_A \rangle &=& \langle  T J_L J_L \rangle + \langle  T J_R
J_R \rangle - \langle  T J_L J_R \rangle - \langle  T J_R J_L \rangle,
\eea

\bea
\langle  T J_A J_A \rangle = \langle  T J_V J_V \rangle - 2 \left( \langle
 T J_L J_R \rangle + \langle  T J_R J_L \rangle \right).
\eea

\beq
\langle T J_L J_L \rangle=\langle T J_R J_R \rangle=\frac{1}{4}\left(\langle T JJ\rangle +\langle T J_A J_A \rangle \right),
\eeq

while
\beq
\langle T_L J_L J_L \rangle= \langle T_R J_R J_R\rangle=  \frac{1}{2} \langle T JJ\rangle
\eeq
is valid for a vanishing fermion mass $m$. The formulation in terms of L/R currents is the most convenient for the study of vertices containing trace anomalies, in the case of realistic theories such as the Standard Model.

\section{Trace anomaly of the $\langle TJ_AJ_A \rangle$ correlator}
We now move to analyze the trace of the $\langle TJ_AJ_A \rangle$ correlator. We consider generic virtualities of the external lines and a massive fermion.

In the absence of anomalies, the naive trace of the $\Gamma_{AA}^{\mu\nu\alpha\beta}$ amplitude is simply obtained by replacing the partial energy-momentum tensor $T_{p}^{\mu\nu}$ in the $\langle TJ_AJ_A \rangle$ correlator with its classical trace $T_{p\,\mu}^{\mu} = - m \bar \psi \psi$ and it is given by
\bea
\Lambda_{AA}^{\alpha \beta}(p,q) &=& -m \, (i g)^2 \int d^4x \, d^4y \, e^{i p\cdot x+iq \cdot y} \langle \bar \psi \psi J_A^{\alpha}(x) J_A^{\beta}(y) \rangle \nn \\
&=& - m \, g^2 \,  \int \frac{d^4 l}{(2 \pi)^4} \, tr \left\{\frac{i}{\lsl-\qsl - m} \g^{\beta}\g_5 \frac{i}{\lsl-m} \g^{\alpha}\g_5 \frac{i}{\lsl+\psl-m} \right\} + \textrm{exch.}
\eea
As in eq.~(\ref{GammaSplit}) we can split the $\Lambda_{A\,A}^{\alpha\beta}$ into two terms: the first, $\Lambda_{VV}^{\alpha\beta}$, being the classical trace obtained from the $\langle TJ_VJ_V \rangle$ correlator, whereas the second, $\Lambda_{\Omega}^{\alpha\beta}$, takes into account the axial contribution to the amplitude as
\bea
\Lambda_{AA}^{\alpha\beta}(p,q) = \Lambda_{VV}^{\alpha\beta}(p,q) + \Lambda_{\Omega}^{\alpha\beta}(p,q).
\label{classicaltrace}
\eea
The $\Lambda^{\alpha\beta}_{VV}$ amplitude refers to the $\langle TJ_VJ_V \rangle$ correlator. It can be written in the form
\bea
\Lambda_{VV}^{\alpha \beta}(p,q) = G_1(s,s_1,s_2,m^2) \, u^{\alpha \beta}(p,q) + G_2(s,s_1,s_2,m^2) \, w^{\alpha \beta}(p,q),
\label{LambdaVV}
\eea
where the rank-2 tensors are defined by
\bea
&&u^{\alpha\beta}(p,q) \equiv (p\cdot q) \,  g^{\alpha\beta} - q^{\alpha} \, p^{\beta}\,,\\
&&w^{\alpha\beta}(p,q) \equiv p^2 \, q^2 \, g^{\alpha\beta} + (p\cdot q) \, p^{\alpha} \, q^{\beta}
- q^2 \,  p^{\alpha} \, p^{\beta} - p^2 \, q^{\alpha} \, q^{\beta},
\label{uwdef}
\eea
with coefficients $G_i(s,s_1,s_2,m^2)$ which are left to an Appendix (Appendix \ref{InvAmp}).\\
The second term $\Lambda_{\Omega}^{\alpha\beta}$ in eq.(\ref{classicaltrace}) can be decomposed into two tensorial structures as
\bea
\Lambda_{\Omega}^{\alpha\beta}(p,q) = H_1(s,s_1,s_2,m^2)g^{\alpha\beta} + H_2(s,s_1,s_2,m^2)(p^{\alpha}q^{\beta}- q^{\alpha}p^{\beta})
\eea
where the functions $H_i$ are related to the invariant amplitudes $R_i$ listed in Appendix \ref{RFormFactors} by the relations
\bea
3 s R_1(s,s_1,s_2,m^2) = H_1(s,s_1,s_2,m^2) - \frac{g^2 m^2}{\pi^2}\,, \\
2 R_2(s,s_1,s_2,m^2) + 2 R_3(s,s_1,s_2,m^2) = H_2(s,s_1,s_2,m^2)\,.
\eea
The analytical expressions of the off-shell $H_i(s,s_1,s_2,m^2)$ form factors are given by
\bea
H_1(s,s_1,s_2,m^2) &=& \frac{g^2\,m^2}{2\pi^2}\bigg[\mathcal D_1(s,s_1,m^2)+\mathcal D_2(s,s_2,m^2)- 2\mathcal B_0(s^2,m^2) +(s-4 m^2)\mathcal C_0(s,s_1,s_2,m^2) \bigg], \nn \\
H_2(s,s_1,s_2,m^2) &=& \frac{g^2 m^2}{\pi^2 \sigma}\bigg[(s+s_1-s_2)\mathcal D_1(s,s_1,m^2)+ (s-s_1+s_2)\mathcal D_2(s,s_2,m^2) \nn \\
&+& s(s-s_1-s_2)\mathcal C_0(s,s_1,s_2,m^2)\bigg],
\eea
where $\si \equiv s^2 - 2 (s_1+s_2)\, s + (s_1-s_2)^2$ and the scalar integrals $\mathcal B_0(s^2,m^2)$, $\mathcal D_1 (s,s_1,m^2)$, $ \mathcal D_2 (s,s_1,m^2)$, $ \mathcal C_0 (s,s_1,s_2,m^2)$ for generic virtualities and masses are defined in Appendix \ref{scalars}.

Tracing the $\Gamma_{AA}^{\mu\nu\alpha\beta}$  correlator we obtain the relation

\bea
g_{\mu \nu}\Gamma_{AA}^{\mu\nu\alpha\beta}(p,q) = \Lambda_{AA}^{\alpha\beta}(p,q) - \frac{g^2}{6\pi^2}u^{\alpha\beta}(p,q)-\frac{g^2 m^2}{\pi^2}g^{\alpha\beta}, \label{GammaTrace}
\eea
where the first term on the right-hand-sice is the trace anomaly appearing already in the $\langle TJ_VJ_V \rangle$ correlator. The second term, proportional to $m^2$, comes from the axial extra term $\Omega^{\mu\nu\alpha\beta}$ and denotes an additional explicit breaking related to the fermion mass. 
In particular, the anomaly $-\frac{g^2}{6\pi^2}u^{\alpha\beta}$ is carried by the form factor $F_1$, whose expression is given in \cite{Armillis:2009pq}, whereas the mass correction $-{g^2 m^2}/{\pi^2} g^{\alpha\beta}$ is induced by $R_1$. This additional contribution 
is gauge variant and its origin can be traced back to the breaking of the $U(1)_A$ gauge symmetry due to the fermion mass term. \\
In the conformal limit the anomalous trace equation (\ref{GammaTrace}) takes a simpler form because, as we have already discussed in the previous sections, the $\langle TJ_AJ_A \rangle$ correlator reduces to the $\langle TJ_VJ_V \rangle$ and we obtain
\begin{figure}[t]
\begin{center}
\includegraphics[scale=1.2]{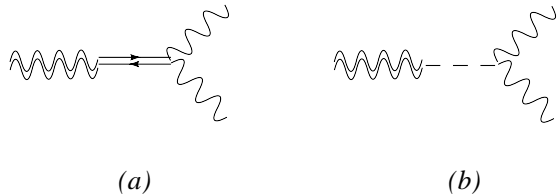}
\caption{\small Polar form of the correlator for external on-shell lines: (a) the contribution to the spectral density from the
collinear on-shell region of the anomaly loop; (b) the pole as virtual exchange in $\Gamma^{anom}$.
 }
\label{collinear}
\end{center}
\end{figure}
%
\bea
g_{\mu \nu}\Gamma_{AA}^{\mu\nu\alpha\beta}(p,q)\bigg\vert_{m=0} = g_{\mu \nu}\Gamma_{VV}^{\mu\nu\alpha\beta}(p,q)\bigg\vert_{m=0} = - \frac{g^2}{6\pi^2}u^{\alpha\beta}(p,q).
\label{trace}
\eea
We give in Appendix  \ref{RFormFactors} the general expression of the form factors $R_i$ ($i=1, \dots, 9$), which, combined with the results of the 13 form factors $F_j$, characterize completely the contributions to the effective action of a vector/axial-vector abelian theory mediated by the conformal anomaly.

Concerning the connection between the anomalous contribution and the $\beta$ function of the theory, also in this case remain valid our previous conclusions, given in \cite{ Giannotti:2008cv,Armillis:2009pq}. Specifically, we just recall, at this point, that
in the (mass independent) regularization scheme $\overline{MS}$ scheme, the $e^2$ term in the trace is directly related to the 
$\beta$ function in this scheme since $\beta(g) ={g^3}/{(12 \pi^2)}$.  In particular, the form factor $F_{13}$ is affected by renormalization via the electric charge \cite{Giannotti:2008cv} \cite{Armillis:2009pq}. 

We close this section with few remarks concerning the structure of the effective action for these types of theories, which can be 
identified from the variational integration of the anomaly equation \cite{Riegert:1984kt}. This approach is, in a way, complementary to the strategy that we follow, based on a direct computation. As shown in \cite{Giannotti:2008cv} there is perfect agreement between the operatorial structure of variational solution, which also exhibits a $1/\square$ effective interaction, and the anomaly pole found in our analysis. In the variational solution of \cite{Riegert:1984kt}, the $1/s$ massless exchange appears after a linearization of the same solution around the flat spacetime limit, as pointed out in \cite{Giannotti:2008cv}. In fact, one obtains in the weak gravitational field limit
\beq
S_{anom}[g,A] = -\frac{c}{6}\int d^4x\sqrt{-g}\int d^4x'\sqrt{-g'}\, R^{(1)}_x
\, \square^{-1}_{x,x'}\, [F_{\alpha\beta}F^{\alpha\beta}]_{x'}\,,
\label{simplifies}
 \eeq
 $(c=-g^2/(24 \pi^2))$. In this case
 \beq
 R^{(1)}_x\equiv \partial^x_\mu\, \partial^x_\nu \, h^{\mu\nu} - \square \,  h, \qquad h=\eta_{\mu\nu} \, h^{\mu\nu}
 \eeq
is the linearized scalar curvature. As in the case of the TJJ correlator \cite{Giannotti:2008cv} the anomalous contribution
to the trace is all contained in the (conformal) anomaly pole (Fig.~\ref{collinear} b)
\beq
\Gamma_{anom}^{\mu\nu\alpha\beta}(p,q) = \int\,d^4x\,\int\,d^4y\, e^{ip\cdot x + i q\cdot y}\,\frac{\delta^2 T^{\mu\nu}_{anom}(0)}
{\delta A_{\alpha}(x) \delta A_{\beta}(y)} = \frac{g^2}{18\pi^2} \frac{1}{k^2} \left(g^{\mu\nu}k^2 - k^{\mu}k^{\nu}\right)u^{\alpha\beta}(p,q)\,,
\label{anom}
\eeq
where  \cite{Giannotti:2008cv}
\beq
T^{\mu\nu}_{anom}(z) =
\frac{c}{3}  \left(g^{\mu\nu}\square - \partial^{\mu}\partial^{\nu}\right)_z \int\,d^4x'\, \square_{z,x'}^{-1}
\left[F_{\alpha\beta}F^{\alpha\beta}\right]_{x'}.
\label{polecontrib}
\eeq

This effective action is trivially obtained from the tensor structure $F_1 t_1^{\mu\nu\alpha\beta}$, present in the expansion of $\Gamma^{\mu\nu\alpha\beta}$ and accounts for the full trace of the correlator in the massless fermion limit, as shown in Eq. (\ref{trace}). 

\subsection{Infrared couplings of the anomaly poles and UV behaviour}
Before coming to conclusions, we pause here in order to comment on these results and on their meaning on a wider perspective. 

We recall that a similar analysis in the QED case  \cite{Giannotti:2008cv, Armillis:2009pq} also manifests such pole singularities, which appear to be rather generic in anomaly amplitudes. They can be attributed, diagrammatically, to specific configurations  of the loop momenta, as illustrated in Fig. (\ref{collinear}). The diagram in this figure describes a massive external line decaying into two massless intermediate fermions, in turn decaying into two on-shell 
axial (or vector) lines (the equivalence between the axial and the vector case in the massless limit is the content of Eq. 
\ref{GammaSplit} ($\Omega\to 0$)). 

The pole is detected by a computation of the spectral density ($\rho(s)$),  which turns out to be proportional to a delta-function
$(\rho(s)\sim \delta(s))$. $\rho(s)$ can be found just by evaluating the $s$-channel cut of the anomalous graph using Cutkovsky rules. This approach, as discussed before \cite{Giannotti:2008cv, Armillis:2009pq}, allows to identify the anomaly poles which are of infrared origin ($s\sim 0$). Other contributions, also characterized by form factors of the form $1/s$, as we have shown, appear in the anomalous amplitude when one performs an {\em off-shell} computation of the anomalous correlator.  These contributions describe the UV behaviour of an anomalous amplitude ($s\to \infty$) and as such they are usually referred to as ``ultraviolet poles", although the name is slightly misleading, being only generated after an asymptotic expansion of the massive correlator. In fact, the residue of the correlator as $s\to 0$ is indeed vanishing in the massive fermion case $\cite{Armillis:2009pq}$, showing that no pole is coupled in this limit. Apart from this important detail, it is however correct to retain their appearance in a perturbative computation - even in the UV region - as a manifestation of the same phenomenon of the trace anomaly. In the case of the chiral anomaly the situation is identical.

These computations \cite{Armillis:2009pq} show that the asymptotic expansion - at large energy - of the regulated graphs responsible for the trace anomaly can be accompanied by corrections which are suppressed as $m^2/s^2$ (as $s \gg m^2$) in the high energy limit, where $m$ is the mass of the fermion in the virtual loop. This organization of the effective action in the UV region allows to recover the ordinary radiative breaking of scale invariance at high energy, being mass corrections negligible in this regime. The use of a mass-independent regularization scheme, such as DRED or DR, is perfectly well taylored in this case, since the separation between pole term and mass corrections involves an asymptotic expansion (at high energy). In particular the $\beta$ function computed in such 
schemes consistently accounts for the UV running of the coupling \cite{Armillis:2009pq}.

We have described this point at length in the case of the gauge anomaly in \cite{Armillis:2009sm}, to which we refer for more details. This implies that the anomaly is saturated by a pole in very different kinematical regions, in agreement with previous analysis performed in chiral theories \cite{Armillis:2009sm, Knecht:2003xy}. 

These conclusions show that the description of the effective action in terms of two auxiliary fields - which are introduced in order to recover the local form of the Lagrangian - is significant both in massless theories \cite{Giannotti:2008cv, Mottola:2006ew} (for instance on null surfaces, i.e. $s=0$), but also in the high energy domain, for large values of $s$. We refer to  \cite{Giannotti:2008cv, Mottola:2006ew} for a discussion of the auxiliary field formulation. 
Similar arguments have been presented in \cite{Coriano:2008pg,Armillis:2009sm, Armillis:2008bg} for the axion pole in the chiral coupling of anomalous $U(1)$'s (in the $AVV$ vertex), proving that these auxiliary degrees of freedom are the most significant signature of chiral and conformal anomalies.

\section{Conclusions}

We have presented an off-shell computation of the correlator of the energy momentum tensor and two vector/axial-vector currents in a chiral theory with an anomalous fermion spectrum, useful for the study of the coupling of anomalous $U(1)$'s to gravity. These interactions are mediated by the trace anomaly.
Starting directly from the functional integral, we have derived the Ward identities for the corresponding vertices. These apply, in general, to any correlator of similar type. All the computations have been performed using DRED, and we have shown the cancellation of mixed chiral/conformal anomalies for these types of vertices. 

Our computation can be viewed as the generalization of the classical analysis of the $AVV$ diagram to these new vertices. We have allowed explicit mass breaking terms to investigate the most general form of the Ward identities for these correlators, that we anticipate of being of crucial importance for the more general analysis in the Standard Model case.

Obviously, the inclusion of our current study into a theory with spontaneous symmetry breaking and Yukawa couplings, such as the Standard Model, would allow to relate the explicit chiral symmetry breaking terms (mass terms) to the extra interactions of the theory,
in particular to the Higgs sector. 

We have also shown that, similarly to the case of a vector-like theory, also in the case of a mixed vector/axial-vector theory, the effective action obtained by coupling gravity to the gauge currents is characterized by effective massless degrees of freedom. A more general analysis of these issues and, in particular, an application of the methods developed in this work in the analysis of anomaly mediation 
in the Standard Model will be presented in a related work.

\centerline{\bf Acknowledgements}
We thank Emil Mottola for discussions during recent visits at CERN. This work is supported in part  by the European Union through the Marie Curie Research and Training Network Universenet (MRTN-CT-2006-035863).

\begin{appendix}
\section{Appendix. Definitions and conventions for the scalar integrals}
\label{scalars}
We collect here the expressions of the three master integrals which appear in the computation in order to be self-contained.
The one-, two- and three-point functions are respectively given by
\bea
\mathcal A_0 (m^2) &=& \frac{1}{i \pi^2}\int d^n l \, \frac{1}{l^2 - m^2}
= m^2 \left [ \frac{1}{\bar \eps} + 1 - \log \left( \frac{m^2}{\mu^2} \right )\right],\\
 \mathcal B_0 (k^2, m^2) &=&  \frac{1}{i \pi^2} \int d^n l \, \frac{1}{(l^2 - m^2) \, ((l - k )^2 - m^2 )} \nn \\
 &=& \frac{1}{\bar \eps} + 2 - \log \left( \frac{m^2}{\mu^2} \right ) - a_3 \log \left( \frac{a_3+1}{a_3-1}\right), \\
\mathcal C_0 (s, s_1, s_2, m^2) &=&
 \frac{1}{i \pi^2} \int d^n l \, \frac{1}{(l^2 - m^2) \, ((l -q )^2 - m^2 ) \, ((l + p )^2 - m^2 )} \nn \\
&=&- \frac{1}{ \sqrt \sigma} \sum_{i=1}^3 \left[Li_2 \frac{b_i -1}{a_i + b_i}   - Li_2 \frac{- b_i -1}{a_i - b_i} + Li_2 \frac{-b_i +1}{a_i - b_i}  - Li_2 \frac{b_i +1}{a_i + b_i}
   \right],
\label{C0polylog}
\eea
with $\si \equiv s^2 - 2 (s_1+s_2)\, s + (s_1-s_2)^2$, 
\bea
\frac{1}{\bar \epsilon} = \frac{1}{\epsilon} -\gamma - \log\pi, \qquad \qquad
a_i = \sqrt {1- \frac{4 m^2}{s_i }}, \qquad \qquad
b_i = \frac{- s_i + s_j + s_k }{\sqrt{ \sigma}},
\eea
where $s_3=s$ and in the last equation $i=1,2,3$ and $j, k\neq i$. \\

We organize the perturbative expansion in terms of two finite combinations of scalar functions given by
\bea
&&  \mathcal B_0 (s, m^2) \, m^2 - \mathcal A_0 (m^2) =  m^2 \left[ 1 - a_3 \log \frac{a_3 +1}{a_3 - 1}  \right] , \\
&& \mathcal D_i \equiv \mathcal D_i (s, s_i,  m^2) =
\mathcal B_0 (s, m^2) - \mathcal B_0 (s_i, m^2) =  \left[ a_i \log\frac{a_i +1}{a_i - 1}
- a_3 \log \frac{a_3 +1}{a_3 - 1}  \right] \qquad i=1,2.
\label{D_i}
\nn \\
\eea
\section{Appendix. Form factors for the off-shell $\langle TJ_AJ_A \rangle$ correlator}
\label{RFormFactors}
This appendix contains the form factors involved in the decomposition of the $\langle TJ_AJ_A \rangle$ correlator, as in eq.(\ref{OmegaR}), expressed in terms of scalar integrals after the tensorial reduction
\begin{align}
R_1(s,s_1,s_2,m^2) &=
 \frac{g^2 \, m^2}{6 \pi^2 \, s} \bigg[\mathcal D_1(s,s_1,m^2)+\mathcal D_2(s,s_2,m^2)- 2 \mathcal B_0(s^2,m^2) - 2\nn \\
 &+ (s-4 m^2)\mathcal C_0(s,s_1,s_2,m^2) \bigg]\\
R_2(s,s_1,s_2,m^2) &= \frac{g^2\,m^2}{4\pi^2\,\sigma} \bigg[ 2 (s-s_1-s_2)\mathcal D_1(s,s_1,m^2) + 4 s_2\mathcal D_2(s,s_2,m^2) \nn \\
&+ ((s-s_1)^2-s_2^2) \mathcal C_0(s,s_1,s_2,m^2) \bigg]\\
R_3(s,s_1,s_2,m^2) &= \frac{g^2\,m^2}{4\pi^2\,\sigma} \bigg[ 4 s_1\mathcal D_1(s,s_1,m^2) + 2 (s-s_1-s_2) \mathcal D_2(s,s_2,m^2) \nn \\
&+ ((s-s_2)^2-s_1^2)\mathcal C_0(s,s_1,s_2,m^2) \bigg]\nonumber\\
R_4(s,s_1,s_2,m^2) &= R_2(s,s_1,s_2,m^2) \\
R_5(s,s_1,s_2,m^2) &= R_3(s,s_1,s_2,m^2) \\
R_6(s,s_1,s_2,m^2) &= \frac{g^2 m^2}{2 \pi^2 (s-s_1-s_2)}\bigg[2\mathcal B_0(s,m^2) - \mathcal D_1(s,s_1,m^2)-\mathcal D_2(s,s_2,m^2)\bigg] \\
R_7(s,s_1,s_2,m^2) &=
\frac{g^2 m^2}{24 \pi ^2 s} \bigg[2\mathcal B_0(s,m^2)+\frac{2}{\sigma^2}\mathcal C_0(s,s_1,s_2,m^2)
   \left(\left(s-s_1\right){}^2+s_2^2+4 s s_2-2 s_1 s_2\right) \times \nn \\
   & \left(2 m^2
   \left(s^2-2 \left(s_1+s_2\right) s+\left(s_1-s_2\right){}^2\right)+s
   \left(s^2-2 \left(s_1+s_2\right) s+s_1^2+s_2^2+4 s_1
   s_2\right)\right)\nn \\
&+ \frac{\mathcal D_1(s,s_1,m^2)}{\sigma^2} \bigg(5 s^4-2
   \left(7 s_1+s_2\right) s^3+4 \left(3 s_1^2+5 s_2 s_1-3 s_2^2\right) s^2 \nn \\
&- 2 \left(s_1-s_2\right) \left(s_1^2+12 s_2 s_1+5 s_2^2\right)
   s-\left(s_1-s_2\right){}^4\bigg) \nn \\
&+\frac{\mathcal D_2(s,s_2,m^2)}{\sigma^2} \bigg(-2 \left(9
   s^2+8 s_1 s+3 s_1^2\right) s_2^2-\left(s-s_1\right){}^4+4 \left(7
   s+s_1\right) s_2 \left(s-s_1\right){}^2 \nn \\
&- s_2^4+4 \left(s_1-2 s\right) s_2^3\bigg)+\frac{2
   \left(\left(s-s_1\right){}^2+s_2^2+4 s s_2-2 s_1 s_2\right)}{\sigma}\bigg]  \displaybreak[0]\\
%
R_8(s,s_1,s_2,m^2) &= \frac{g^2 m^2}{24 \pi ^2 s} \bigg[2 \mathcal B_0(s,m^2)+\frac{2 \mathcal C_0(s,s_1,s_2,m^2)}{\sigma^2} \left(s^2+4 s_1
   s+s_1^2+s_2^2-2 \left(s+s_1\right) s_2\right)\times \nn \\
& \left(2 m^2 \left(s^2-2
   \left(s_1+s_2\right) s+\left(s_1-s_2\right){}^2\right)+s \left(s^2-2
   \left(s_1+s_2\right) s+s_1^2+s_2^2+4 s_1 s_2\right)\right)  \nn \\
&- \frac{\mathcal D_1(s,s_1,m^2)}{\sigma^2} \bigg(s^4-4 \left(7
   s_1+s_2\right) s^3+2 \left(9 s_1^2+26 s_2 s_1+3 s_2^2\right) s^2 \nn \\
&+ 4 \left(s_1-s_2\right) \left(2 s_1^2+6 s_2 s_1+s_2^2\right)
   s+\left(s_1-s_2\right){}^4\bigg) \nn \\
&+ \frac{\mathcal D_2(s,s_2,m^2)}{\sigma^2} \bigg(5 s^4-2
   \left(s_1+7 s_2\right) s^3+4 \left(-3 s_1^2+5 s_2 s_1+3 s_2^2\right) s^2 -\left(s_1-s_2\right){}^4 \nn \\
&+ 2 \left(s_1-s_2\right) \left(5 s_1^2+12 s_2 s_1+s_2^2\right)
   s\bigg)+\frac{2 \left(s^2+4 s_1
   s+s_1^2+s_2^2-2 \left(s+s_1\right) s_2\right)}{\sigma}\bigg] \nn
   \displaybreak[0]\\ \\
R_9(s,s_1,s_2,m^2) &= \frac{g^2 m^2}{12 \pi ^2 s} \bigg[2 \mathcal B_0(s,m^2)\left(1+\frac{3 s}{\gamma}\right)-\frac{2 \mathcal C_0(s,s_1,s_2,m^2)}{\sigma^2} \left(2
   s^2-\left(s_1+s_2\right) s-\left(s_1-s_2\right){}^2\right) \times \nn \\
& \left(2 m^2
   \left(s^2-2 \left(s_1+s_2\right) s+\left(s_1-s_2\right){}^2\right)+s
   \left(s^2-2 \left(s_1+s_2\right) s+s_1^2+s_2^2+4 s_1
   s_2\right)\right) \nn \\
&+ \frac{\mathcal D_1(s,s_1,m^2)}{\gamma \, \sigma^2} \bigg(-10
   s^5+\left(23 s_1+41 s_2\right) s^4-2 \left(5 s_1^2+27 s_2 s_1+32
   s_2^2\right) s^3 \nn \\
&- 2 \left(s_1+s_2\right) \left(4 s_1^2+5 s_2 s_1-23
   s_2^2\right) s^2+2 \left(s_1-s_2\right) \left(2 s_1^3+19 s_2 s_1^2+8 s_2^2
   s_1+7 s_2^3\right) s \nn \\
&+ \left(s_1-s_2\right){}^4
   \left(s_1+s_2\right)\bigg) \nn \\
&+ \frac{\mathcal D_2(s,s_2,m^2)}{\gamma \, \sigma^2} \bigg(2 \left(-4
   s^2+17 s_1 s+s_1^2\right) s_2^3+\left(s-s_1\right){}^2 \left(23 s^2-8 s_1
   s-3 s_1^2\right) s_2 \nn \\
&- 2 \left(5 s^3+9 s_1 s^2+11 s_1^2 s-s_1^3\right)
   s_2^2+s_2^5+\left(4 s-3 s_1\right) s_2^4-\left(s-s_1\right){}^4 \left(10
   s-s_1\right)\bigg) \nn \\
&+ \frac{2
   \left(-2 s^2+\left(s_1+s_2\right) s+\left(s_1-s_2\right){}^2\right)}{\sigma}\bigg],
\end{align}
where $s=k^2=(p+q)^2$, $s_1=p^2$, $s_2=q^2$, $\gamma \equiv s -s_1 - s_2$, $\si \equiv s^2 - 2 (s_1+s_2)\, s + (s_1-s_2)^2$ and the scalar integrals $\mathcal B_0(s^2,m^2)$, $\mathcal D_1 (s,s_1,m^2)$, $ \mathcal D_2 (s,s_1,m^2)$, $ \mathcal C_0 (s,s_1,s_2,m^2)$ for generic virtualities and masses are defined in Appendix \ref{scalars}.

\section{Appendix. Form factors for the $\Lambda_{VV}^{\alpha\beta}$ amplitude}
\label{InvAmp}
We write in this appendix the form factors $G_1$ and $G_2$ appearing in eq.~\ref{LambdaVV} as contributions to the classical trace obtained for the $\langle T J_V J_V\rangle$ correlator
\bea
G_1(s,s_1,s_2,m^2) &=&
 \frac{ g ^2 \gamma   m^2}{ \pi^2 \sigma } +\frac{g^2\, \mathcal D_2(s,s_2,m^2)\,  s_2 m^2}{ \pi^2 \sigma ^2}
\left[s^2+4 s_1 s-2 s_2 s-5 s_1^2+s_2^2+4 s_1   s_2\right]    \nn \\
&&   \hspace{-3cm}
-   \frac{g^2 \, \mathcal D_1 (s,s_1,m^2)\, s_1 m^2}{ \pi^2 \sigma ^2}
   \left[-\left(s-s_1\right){}^2+5 s_2^2-4 \left(s+s_1\right)
   s_2\right]   \nn \\
  && \hspace{-3cm}
 - g^2 \, \mathcal C_0 (s,s_1,s_2,m^2)\,
 \left[
\frac{ m^2 \gamma}{2 \pi^2  \sigma^2}   \left[ \left(s-s_1\right){}^3-s_2^3+\left(3 s+s_1\right)  s_2^2+\left(-3 s^2-10 s_1 s+s_1^2\right) s_2 \right] -\frac{2 m^4 \gamma }{ \pi^2 \sigma }\right], \nn \\ \\
G_2(s,s_1,s_2,m^2) &=&
 - \frac{2 g^2 m^2}{ \pi^2 \sigma }
 - \frac{2 g^2 \, \mathcal D_2(s,s_2,m^2) m^2}{\pi^2 \sigma   ^2} \,  \left[\left(s-s_1\right){}^2-2
   s_2^2+\left(s+s_1\right) s_2\right]       \nn \\
   && \hspace{-3cm}
   -\, \frac{2 \, g^2 \mathcal D_1 (s,s_1,m^2) \, m^2}{\pi^2 \sigma ^2}
   \left[s^2+\left(s_1-2 s_2\right) s-2 s_1^2+s_2^2+s_1 s_2\right] \nn \\
    && \hspace{-3cm}
 - \, g^2 \mathcal C_0 (s,s_1,s_2,m^2) \,
    \biggl[\frac{4 m^4}{ \pi^2 \sigma}
  +\frac{m^2}{\pi^2 \sigma ^2} \,  \left[ s^3-\left(s_1+s_2\right) s^2
  - \left(s_1^2-6  s_2 s_1+s_2^2\right) \right. s \nn \\
  && \hspace{3 cm} +\left. \left(s_1-s_2\right){}^2
   \left(s_1+s_2\right)\right] \biggr],
\eea
where $\gamma \equiv s -s_1 - s_2$, $\si \equiv s^2 - 2 (s_1+s_2)\, s + (s_1-s_2)^2$ and the scalar integrals $ \mathcal D_1 (s,s_1,m^2)$, $ \mathcal D_2 (s,s_1,m^2)$ and $ \mathcal C_0 (s,s_1,s_2,m^2)$ have been already defined in Appendix \ref{scalars}.

\end{appendix}


\end{document}